\documentclass[
 preprint,
nofootinbib,
 amsmath,amssymb,
 aps,
]{revtex4}
\usepackage{graphicx}% Include figure files
\usepackage{dcolumn}% Align table columns on decimal point
\usepackage{bm}% bold math
\usepackage{hyperref}% add hypertext capabilities
\usepackage[normalem]{ulem}
\usepackage{color}
\usepackage{caption}
\usepackage{subcaption}

\begin{document}

\title{{Non-zero $\theta_{13}$, CP-violation and Neutrinoless Double Beta Decay for Neutrino Mixing in the $A_4\times Z_2\times Z_3$ Flavor Symmetry Model}}%

\author{Animesh Barman}
    \thanks{Email address: animesh@tezu.ernet.in  (Corresponding author)}
\author{Ng. K. Francis}
    \thanks{Email address: francis@tezu.ernet.in}
\author{Bikash Thapa}
    \thanks{Email address: bikash2@tezu.ernet.in}
\author{Ankur Nath}
    \thanks{Email address: ankur04@tezu.ernet.in}
    
\affiliation{Department of Physics, Tezpur University, Tezpur - 784028, India}

%\date{\today}% It is always \today, today,
             %  but any date may be explicitly specified
             
\begin{abstract}
    We study the modification of the Altarelli-Feruglio $A_4$ flavor symmetry model by adding three singlet flavons $\xi'$, $\xi''$ and $\rho$ and the model is augmented with extra {$Z_2\times Z_3$} symmetry to prevent the unwanted terms in our study. The addition of these three  flavons lead to two higher order corrections in the form of two perturbation parameters $\epsilon$ and $\epsilon^\prime$. These corrections yield the deviation from exact tri-bimaximal (TBM) neutrino mixing pattern by producing a non-zero $\theta_{13}$ and other neutrino oscillation parameters which are consistent with the latest experimental data. In both the corrections, the neutrino masses are generated via Weinberg operator. The analysis of the perturbation parameters $\epsilon$ and $\epsilon^\prime$, shows that normal hierarchy (NH) and inverted hierarchy (IH) for $\epsilon$ does not change much. However, as the values of $\epsilon^\prime$ increases, $\theta_{23}$ occupies the lower octant for NH case. We further investigate the neutrinoless double beta decay parmeter $m_{\beta\beta}$ using the parameter space of the model for both normal and inverted hierarchies of neutrino masses.\\
    
\end{abstract}             
%\keywords{Suggested keywords}%Use showkeys class option if keyword
                              %display desired

\maketitle

%\tableofcontents

\section{\label{sec:1}Introduction}

Though the particle physics experiments and observations have been successfully confirming the standard model (SM) of particle physics,  the origin of flavor structure, strong CP problem, matter-antimatter asymmetry of the universe, dark matter, dark energy, non-zero tiny neutrino masses, presence of extra flavor of neutrinos, etc. are still open questions. The discovery of neutrino oscillations in 1998 by Super-Kamiokande (SK), Japan and Sudbury Neutrino Observatory (SNO), Canada are the first proof of physics beyond the standard model. In  neutrino physics, we still do not know if the leptonic CP symmetry is violated or not, whether the neutrino mass eigenvalues follow normal hierarchy (NH) or inverted hierarchy (IH), if the atmospheric mixing angle $\theta_{23}$ is maximal or not,  are the neutrinos of the Dirac or Majorana type, the absolute mass of the lightest  neutrino flavor etc. The neutrino oscillation experiments are only sensitive to the mass squared differences $\Delta m_{ij}^2$, and the leptonic mixing angles $\theta_{ij}$ ($i=1,2,3$).

Neutrino physics is an experimentally driven  field. It has made tremendous progress over the past few decades and attempts are underway to quantify the neutrino oscillation parameters more precisely. A few  notable works in neutrino physics are placed in references \cite{King:2014nza, King:2003jb, Cao:2020ans,  McDonald:2016ixn, Kajita:2016cak, Vien:2020dlk, deSalas:2020pgw, King:2017guk, PhongNguyen:2017meq, Nguyen:2020ehj}.

Neutrino oscillation phenomenology is characterized by two large mixing angles, the solar angle $\theta_{12}$ and the atmospheric angle $\theta_{23}$ together with the relatively small reactor mixing angle $\theta_{13}$. In tri-bimaximal mixing (TBM), the reactor mixing angle $\theta_{13}$ is zero and the CP phase $\delta_{CP}$ is consequently undefined. However, in 2012 the Daya Bay reactor neutrino experiment ($\sin^2 2\theta_{13} = 0.089 \pm 0.010 \pm 0.005 $) \cite{DayaBay:2012fng} and RENO experiment $\sin^2 2\theta_{13}= 0.113 \pm 0.013 \pm 0.019 $ \cite{RENO:2012mkc} showed that $\theta_{13} \simeq 9 ^\circ$.  Several accelerator-based long baseline neutrino oscillation experiments like MINOS \cite{MINOS:2011amj}, Double Chooz \cite{DoubleChooz:2011ymz}, T2K \cite{T2K:2011ypd} also measured consistent non-zero values for $\theta_{13}$. Since TBM has been ruled out due to a non-zero reactor mixing angle, \cite{RENO:2012mkc, DoubleChooz:2011ymz} one of the admired ways to achieve realistic mixing is through either it's extensions or through modifications.

The widely accepted PMNS matrix encodes the mixing between the neutrino flavor eigen states ($\nu_e$, $\nu_\mu$, $\nu_\tau$) and their mass eigen states ($\nu_1$, $\nu_2$, $\nu_3$). In the three flavored paradigm, three mixing angles and one CP phase are used to parameterize this PMNS matrix , given by Equation~\ref{eq:1}.
\begin{equation}
\label{eq:1}
    U_{PMNS}=
    \begin{pmatrix}
    c_{12} c_{13} & s_{12} c_{13} & s_{13}e^{-i \delta}\\
    -s_{12} c_{23}- c_{12} s_{23} s_{13} e^{ i \delta} & c_{12} c_{23} - s_{12} s_{23} s_{13} e^{i \delta} & s_{23} c_{13}\\
    s_{12} s_{23} - c_{12} c_{23} s_{13} e^{ i \delta} & -c_{12} s_{23} -s_{12} c_{23} s_{13} e^{ i \delta} & \ c_{23} c_{13}
    \end{pmatrix}
    \cdot U_{Maj}
\end{equation}
where, $ c_{ij}=\cos{\theta_{ij}}$, $s_{ij}=\sin{\theta_{ij}}$  such that ($i,j=1,2,3$ and $i<j$). The diagonal matrix $U_{Maj}= diag (1, e^{ i \alpha}, e^{ i\beta})$ contains the Majorana CP phases, $\alpha$ and  $\beta$, which become observable in case the neutrinos behave as Majorana particles. To show that neutrinos are Majorana particles, it  require neutrino-less double beta decay to be discovered. Such kind of decays are yet to be observed. To explain these issues, symmetry would play an important role. Wendell Furry \cite{Furry:1939qr} considered Majorana nature of particles, to study a kinetic process which was similar to double beta disintegration without neutrino emission popularly known as neutrino-less double beta decay (NDBD) \cite{DellOro:2016tmg}. It can be expressed as 
$(A,Z)\rightarrow (A, Z+2)+2e^-$ which violates the lepton number by two units and creates a pair of electron, and Majorana neutrino masses are generated as electroweak symmetry is broken. The large value of the cut-off scale of lepton number violation (LNV), typically $\Lambda \sim (10^{14} - 10^{15})$ GeV, is generally linked to the observed smallness of neutrino masses. Since, the neutrino mass is zero in standard model\cite{Bilenky:2012qi}, we need to construct a model which is beyond the standard model by adopting a new symmetry and generate non-zero neutrino mass. One such model is the effective theories, which can generate neutrino masses through Weinberg operator \cite{Weinberg:1979sa}. 

There are other frameworks beyond the standard model (BSM) that can explain the origin of neutrino masses, for examples, the Seesaw Mechanism \cite{Yanagida:1980xy,Minkowski:1977sc, Das:2019kmn, gell1979ramond, Mohapatra:1979ia,Fukuyama:1997ky,Vien:2019zhs, yanagida1979horizontal, Boruah:2021ktk}, Supersymmetry \cite{Ma:1998ias},  Minimal Supersymmetric Standard Model (MSSM)\cite{Csaki:1996ks}, Next-to-Minimal Supersymmetric Standard Model (NMSSM)\cite{Ellwanger:2009dp}, String theory \cite{Ibanez:2012zz}, models based on extra dimensions \cite{Arkani-Hamed:1998wuz}, Radiative Seesaw Mechanism \cite{ Ma:2006km} and also some other models. Now, many neutrino experiments have proved  beyond doubt that neutrino has tiny non-zero mass and indicate flavor mixing \cite{aker2019improved, Francis:2014dya, particle2020review, Nath:2018ywc}.

Tri-bimaximal (TBM) \cite{Harrison:2002er, Harrison:2002kp},Trimaximal (TM1/TM2) \cite{Albright:2008rp, He:2011gb, thapa2021resonant}, Quasi-degenerate neutrino mass models \cite{Francis:2012jj}  and Bi-large mixing patterns \cite{Boucenna:2012xb, Chen:2019egu, Ding:2019vvi} are examples of phenomenological neutrino mixing patterns. Also, various models based on non-abelian discrete flavor symmetries \cite{King:2013eh} like $A_4$ \cite{Ishimori:2010au, Ma:2005qf, Ma:2015fpa}, $S_3$ \cite{Ma:2004zd}, $S_4$ \cite{Bazzocchi:2012st, Ma:2005mw, Vien:2015fhk}, $\Delta_{27}$ \cite{ma2008near, de2007neutrino, harrison2014deviations}, $\Delta_{54}$ \cite{Ishimori:2008uc,loualidi2021trimaximal} have been proposed to obtain tri-bimaximal mixing (TBM).

Our model is based on the Altarelli-Feruglio (A-F) $A_4$ discrete flavor symmetry model \cite{Altarelli:2005yx,Altarelli:2005yp,Altarelli:2010gt}. We have extended the flavon sector of A-F model by introducing extra flavons $\xi^\prime$, $\xi^{\prime\prime}$ and $\rho$ which transform as $1^\prime$, $1^{\prime\prime}$ and 1 respectively under $A_4$ to get the deviation from exact TBM neutrino mixing pattern. We also introduce $Z_2\times Z_3$ symmetry in our model to prevent unwanted terms and this helps in constructing specific structure of the coupling matrices. We calculate higher dimension perturbative parameters $\epsilon$ and $\epsilon^\prime$ from the Lagrangian, modified the neutrino mass matrix $M_\nu$ and realized symmetry based studies. However, in a few similar works \cite{Brahmachari:2008fn, Shimizu:2011xg}, they have been simply added arbitrarily perturbative terms to $M_\nu$ obtained from A-F model without calculations from the Lagrangian. But in these papers \cite{Karmakar:2014dva, King:2011zj, Cooper:2011rh, Ding:2011gt}, perturbative term was calculated using Type-I seesaw mechanism and departed from the tri-bimaximal mixing pattern. 

Also, various efforts have been made to deviate from TBM structure by adding extra flavons in order to generate non-zero $\theta_{13}$, non-trivial CP phase, thereby explaining the experimental data. A few of them like \cite{Ahn:2012tv} where the author shows that non-degeneracy of the neutrino Yukawa coupling constants are the origin of the deviations from the TBM mixing and unremovable CP-phases in the neutrino Yukawa matrix give rise both low energy CP violation measurable from neutrino oscillation and high energy CP violation. In Ref.~\cite{Ahn:2013mva} the authors show that CP is spontaneously broken at high energies, after breaking of flavon symmetry by a complex vacuum expectation value of $A_4$ triplet and gauge singlet scalar field. In Ref.~\cite{Kang:2018txu}, Dirac CP violating phase is predicted by using the experimental mixing parameters and this model is consistent with the experimental data only for the normal hierarchy of neutrino masses.

Therefore, this gives us an opportunity to analyze $M_\nu$ obtained through Weinberg operator in detail and study the effect of two perturbative terms $\epsilon$ and $\epsilon^\prime$ on neutrino oscillation parameters and NDBD parameter $m_{ee}$. 

The content material of our paper is organised as follows: In section 2, we give the overview of the framework of our model by specifying the fields involved and their transformation properties under the symmetries imposed. We give two types of corrections, analyse  and study the impact of these correction terms on neutrino oscillation parameters. In section 3, we do numerical analysis and study the results for the neutrino phenomenology. We finally conclude our work in section 4.

\begin{table}[t]
\centering
  \begin{tabular}{ | l | c | r |}
    \hline
    Parameters & NH (3$\sigma$) & IH (3$\sigma$) \\ \hline
    $\Delta{m}^{2}_{21}[{10}^{-5}eV^{2}]$ & $6.82 \rightarrow 8.04$ & $6.82 \rightarrow 8.04$ \\ \hline
    $\Delta{m}^{2}_{31}[{10}^{-3}eV^{2}]$ & $2.431 \rightarrow 2.599$ & $-2.584 \rightarrow -2.413 $\\ \hline
    $\sin^{2}\theta_{12}$ & $0.269 \rightarrow 0.343$ & $0.269 \rightarrow 0.343$ \\ \hline
     $\sin^{2}\theta_{13}$ & $0.02034 \rightarrow 0.02430$ & $0.02053 \rightarrow 0.02434$ \\ \hline
    $\sin^{2}\theta_{23}$ & $0.405 \rightarrow 0.620$ & $0.410 \rightarrow 0.623$ \\ \hline
    $\delta_{CP}$ & $105 \rightarrow 405$ & $192 \rightarrow 361$ \\ \hline
  \end{tabular}
    \caption{The $3\sigma$ ranges of neutrino oscillation parameters from NuFIT 5.1 (2021) \cite{Esteban:2020cvm}}
    \label{tab:1}
\end{table}

\section{\label{sec:f}Framework of the Model}
    The non-Abelian discrete symmetry group $A_4$ is a group of even permutations of four objects and it has 12 elements (12= $\frac{4!}{2}$). It can describe the orientation-preserving symmetry of a regular tetrahedron, so this group is also known as tetrahedron group. It can be generated by two basic permutations S and T having properties $S^2=T^3=(ST)^3=1$. This group representations include three one-dimensional unitary representations $1$, $1^\prime$,  $1^{\prime\prime}$ with the generators S and T given, respectively as follows:
$$1: S=1, T=1$$
$$1^\prime: S=1, T=\omega^2$$
$$1^{\prime\prime}: S=1, T=\omega$$

and a three dimensional unitary representation with the generators\footnote{Here the generator T has been chosen to be diagonal.}
\begin{equation}
    T=
    \begin{pmatrix}
    1 & 0 & 0\\
    0 & \omega^2 & 0\\
    0 & 0 & \omega
    \end{pmatrix}
\end{equation}

\begin{equation}
    S=\frac{1}{3}
    \begin{pmatrix}
    -1 & 2 & 2\\
    2 & -1 & 2\\
    2 & 2 & -1
    \end{pmatrix}
\end{equation}\\.
Here, $\omega$ is the cubic root of unity, $\omega=exp(i2\pi)$, so that $1+\omega+\omega^2 =0$.\\
The multiplication rules corresponding to the specific basis of two generators S and T are as follows:
$$1\times1=1$$
$$1^{\prime\prime}\times1^\prime =1$$
$$1^\prime\times\ 1^{\prime\prime}=1$$
$$3\times3=3+ 3_A + 1+ 1^\prime +1^{\prime\prime}$$
For two triplets\\
$$a= (a_1,a_2, a_3)$$
$$b= (b_1,b_2, b_3)$$
we can write
$$1 \equiv (ab) = a_1 b_1 +a_2 b_3 +a_3 b_2$$
$$1^\prime \equiv (ab)^\prime=a_3 b_3 +a_1 b_2 +a_2 b_1$$
$$1^{\prime\prime} \equiv (ab)^{\prime\prime}=a_2 b_2 +a_1 b_3 +a_3 b_1$$

Here, 1 is symmetric under the exchange of second and third elements of a and b, $1^\prime$ is symmetric under the exchange of the first and second elements while $1^{\prime\prime}$ is symmetric under the exchange of first and third elements.
\begin{equation*}
3 \equiv (ab)_S=\frac{1}{3}(2a_1 b_1 -a_2 b_3 -a_3 b_2, 2a_3 b_3 -a_1 b_2 -a_2 b_1, 2a_2 b_2 -a_1 b_3 -a_3 b_1)
\end{equation*} {\begin{align*}
    3_A \equiv (ab)_A =\frac{1}{3}(a_2 b_3 -a_3 b_2, a_1 b_2 -a_2 b_1, a_3 b_1-a_1 b_3 )
\end{align*}}
Here 3 is symmetric and $3_A$ is anti-symmetric. For the symmetric case, we notice that the first element here has 2-3 exchange symmetry, the second element has 1-2 exchange symmetry and the third element has 1-3 exchange symmetry.

\begin{table}[t]
    \centering
    \begin{tabular}{c c c c c c c c c c c c c}
    \hline
          \textrm{Field}  &  $l$ & $e^c$ & $\mu^c$ & $\tau^c$ & $h_u$ & $h_d$ & $\Phi_S$ & $\Phi_T$ & $\xi$ & $\xi^\prime$ & $\xi^{\prime\prime}$ & $\rho$ \\
     \hline
     
     \textrm{SU(2)}  &  2 & 1 & 1 & 1 & 2 & 2 & 1 & 1 & 1 & 1 & 1 & 1 \\
     
     \textrm{A}$_4$  &  3 & 1 & $1^{\prime\prime}$ & $1^\prime$ & 1 & 1 & 3 & 3 & 1 & $1^\prime$ & $1^{\prime\prime}$ & 1 \\
     
    \textrm{Z}$_3$  &  $\omega$ & $\omega^2$ & $\omega^2$ & $\omega^2$ & 1 & 1 & $\omega$ & 1  & $\omega$  & $\omega$  & $\omega$  & 1 \\
    
    \textrm{Z}$_2$  &  l & 1 & 1 & 1 & 1 & 1 & 1 & 1 & 1 & 1 & 1 & -1 \\
    \hline
    \end{tabular}
    \caption{Full particle content of our model.}
    \label{tab:2}
\end{table}

Our model is based on the Altarelli-Feruglio $A_4$ model \cite{Altarelli:2005yx,Altarelli:2005yp,Altarelli:2010gt}. We have added additional flavons $\xi'$, $\xi''$ and $\rho$ to get the deviation from exact TBM neutrino mixing pattern. We put extra symmetry $Z_2\times Z_3$ to avoid unwanted terms. The particle content and their charge assignment under the symmetry group is given in Table \ref{tab:2}. The left-handed lepton doublets and right-handed charged leptons ($e^c, \mu^c, \tau^c$) are assigned to triplet and singlet ($1,  1^{\prime \prime}, 1^\prime $) representation under A$_4$ respectively and other particles transform as shown in Table-II. Here, $h_u$ and $h_d$ are the standard Higgs doublets which remain invariant under $A_4$. There are six $SU(2)\otimes U_Y (1)$ Higgs singlets, four ($\xi$, $\xi^\prime$, $\xi^{\prime\prime}$ and $\rho$) of which singlets under $A_4$ and two ($\Phi_T$ and $\Phi_S$) of which transform as triplets.

Consequently, the invariant Yukawa Lagrangian is as follows:
\begin{equation}
    \mathcal{L} = y_e e^c (\Phi_T l) +y_\mu \mu^c (\Phi_T l)^\prime +y_\tau \tau^c (\Phi_T l)^{\prime\prime} + x_a \xi(l l) +x_a^\prime\xi ^\prime(l l)^{\prime\prime }+ x_a^{\prime\prime} \xi^{\prime\prime} (l l)^\prime + x_b (\Phi_S l l) +h.c.+...
\end{equation}
where we have used the compact notation,
\begin{equation}
y_e e^c (\Phi_Tl) \equiv y_e e^c (\Phi_Tl) \frac{h_d}{\Lambda}
\end{equation}
\begin{equation}
x_a \xi(l l) \equiv x_a \xi \frac{(l h_u l h_u)}{\Lambda^2}
\end{equation}
and so on and $\Lambda$ is the cut-off scale of the theory. The terms $y_e$, $y_\mu$, $y_\tau$, $x_a$, $x_a^\prime$ and $x_a^{\prime\prime}$, $x_b$ are coupling constants. We assume $\Phi_T$ does not couple to the Majorana mass matrix and $\Phi_S$ does not couple to the charged leptons.

After spontaneous symmetry breaking of flavor and electroweak symmetry, we obtain the mass matrices for the charged leptons and neutrinos. The vacuum expectation values (VEV) of the scalar fields are of the form \cite{Altarelli:2005yx,Altarelli:2005yp,Altarelli:2010gt}. For the sake of completeness, we present the explicit form of the scalar potentials and their corresponding VEVs in Appendix~\ref{sec.foo}.\\
$\langle \Phi_T \rangle=(v_T ,0,0)$, $\langle \Phi_S \rangle=(v_S ,v_S, v_S)$, $\langle\xi\rangle=u$,
$\langle\xi^\prime\rangle=u^\prime$, $\langle\xi^{\prime\prime}\rangle =u^{\prime\prime}$,
$\langle h_u\rangle= v_u$ and $\langle h_d\rangle= v_d$.\\
The charged lepton mass matrix\footnote{\{Charged fermion masses are given by~\cite{Altarelli:2005yx}:
$m_e=y_e v_d \frac{V_T}{\Lambda}$,
$m_\mu=y_\nu v_d \frac{V_T}{\Lambda}$,
$m_\tau=y_\tau v_d \frac{V_T}{\Lambda}$.
We can obtain a natural hierarchy among $m_e$, $m_\mu$ and $m_\tau$ by introducing an additional $U(1)_F$ flavor symmetry under which only the right-handed lepton sector is charged. We write the F-charge values in this model as 0, q and 2q for $\tau^c$, $\mu^c$ and $e^c$ respectively. By assuming that a flavon $\theta$, carrying a negative unit of F, acquires a VEV $<\theta> / \Lambda \equiv \lambda<1$, the Yukawa couplings become field dependent quantities 
$ y_{e, \mu, \tau} =y_{e, \mu, \tau} (\theta)$ and we have 
$y_\tau \approx O(1)$, $y_\mu \approx O(\lambda^q)$, $y_e \approx O(\lambda^{2q})$.} is given as 
\begin{equation}
    M_l= \frac{v_d v_T}{\Lambda}
    \begin{pmatrix}
    y_e & 0 & 0\\
    0 & y_\mu & 0\\
    0 & 0 & y_\tau
    \end{pmatrix}
\end{equation}

 where, $v_d$ and $v_T$ are the VEV of $h_d$ and $\Phi_T$ respectively. Now,  taking higher dimension terms in the neutrino sector, we consider two types of corrections of the form $x_\epsilon^{\prime\prime}\xi^{\prime\prime} {(ll)}^{\prime}\frac{\rho\rho}{\Lambda^2}$ and $x_\epsilon^{\prime}\xi^{\prime} {(l l)}^{\prime\prime}\frac{\rho\rho}{\Lambda^2}$, where, $x_\epsilon^\prime$ and $x_\epsilon^{\prime\prime}$ are coupling constants. These give rise to two cases and we will study the impact of these correction terms on neutrino oscillation parameters.
 
 \subsection{Case I}
With the additional higher dimension term $x_\epsilon^{\prime\prime}\xi^{\prime\prime} {(ll)}^{\prime}\frac{\rho\rho}{\Lambda^2}$ and using the VEVs $\langle\Phi_s\rangle=(v_{s},v_{s},v_{s})$, $\langle\xi\rangle=0$,
$\langle\xi^\prime\rangle=u^\prime$, $\langle\xi^{\prime\prime}\rangle =u^{\prime\prime}$,
$\langle h_u\rangle= v_u$ and $\langle \rho\rangle = v_\rho$, we obtain the neutrino mass matrix which may be written as

\begin{equation}
\label{eq:8}
    M^{(I)}_\nu= m_0
    \begin{pmatrix}
     \frac{2b}{3} & c -\frac{b}{3}+ \epsilon & d -\frac{b}{3}\\
    c- \frac{b}{3}+ \epsilon  & d +\frac{2b}{3} &  -\frac{b}{3}\\
    d -\frac{b}{3} & -\frac{b}{3} & c +\frac{2b}{3}+ \epsilon
    \end{pmatrix}
\end{equation}

where, $m_0 =\frac{v_u^2}{\Lambda}$,
$b=2x_b\frac{v_s}{\Lambda}$, $c=2x_a^{\prime\prime}\frac{u^{\prime\prime}}{\Lambda}$, $d=2x_a^\prime\frac{u^\prime}{\Lambda}$ and $\epsilon$=$x_\epsilon^{\prime\prime}\frac{u^{\prime\prime} v_\rho^2}{\Lambda^3}$. 
We can assume c$\simeq$d. This is a reasonable assumption to make since the phenomenology does not change drastically unless the VEVs of the singlet Higgs vary by a huge amount. Thus, the neutrino mass matrix in equation (\ref{eq:8}) becomes

\begin{equation}
\label{eq:9}
    M^{(I)}_\nu= m_0
    \begin{pmatrix}
     \frac{2b}{3} & d -\frac{b}{3}+ \epsilon & d -\frac{b}{3}\\
    d- \frac{b}{3}+ \epsilon  & d +\frac{2b}{3} &  -\frac{b}{3}\\
    d -\frac{b}{3} & -\frac{b}{3} & d +\frac{2b}{3}+ \epsilon
    \end{pmatrix}
\end{equation}

\iffalse

\begin{equation}
   x_\epsilon^{\prime\prime}\xi^{\prime\prime} {(l l)}^{\prime}\frac{\rho\rho}{\Lambda^2}
\end{equation}

which gives rise a correction matrix by

\begin{equation}
    M_\nu^\prime= \frac{v_\rho^2}{\Lambda ^2}u^{\prime\prime}
    \begin{pmatrix}
    0 & 1 & 0\\
    1 & 0 & 0\\
    0 & 0 & 1
    \end{pmatrix}
\end{equation}

\begin{equation}
    M_\nu^\prime= 
    \begin{pmatrix}
    0 & \epsilon & 0\\
    \epsilon & 0 & 0\\
    0 & 0 & \epsilon
    \end{pmatrix}
\end{equation}

where $\epsilon =\frac{v_\rho^2}{\Lambda ^2}u^{\prime\prime}$

Again in another way we put in higher dimension term of the form

\begin{equation}
   x_\epsilon^{\prime}\xi^{\prime} {(l l)}^{\prime\prime}\frac{\rho\rho}{\Lambda^2}
\end{equation}

which gives rise a correction matrix by

\begin{equation}
    M_\nu^{\prime\prime}= \frac{v_\rho^2}{\Lambda ^2}u^{\prime}
    \begin{pmatrix}
    0 & 0 & 1\\
    0 & 1 & 0\\
    1 & 0 & 0
    \end{pmatrix}
\end{equation}

\begin{equation}
    M_\nu^{\prime\prime}= 
    \begin{pmatrix}
    0 & 0 & \epsilon\\
    0 & \epsilon & 0\\
    \epsilon & 0 & 0
    \end{pmatrix}
\end{equation}

where $\epsilon =\frac{v_\rho^2}{\Lambda ^2}u^{\prime}$

\fi

\subsection{Case II}
\label{sec:2}
Here, we will take into consideration the correction term of the second type $x_\epsilon^{\prime}\xi^{\prime} {(l l)}^{\prime\prime}\frac{\rho\rho}{\Lambda^2}$. The resulting neutrino mass matrix obtained in such a case is given as

\begin{equation}
\label{eq:10}
    M^{(II)}_\nu= m_0
    \begin{pmatrix}
    \frac{2b}{3} & c -\frac{b}{3} & d -\frac{b}{3}+ \epsilon^\prime \\
    c- \frac{b}{3}  & d +\frac{2b}{3}+ \epsilon^\prime & -\frac{b}{3}\\
    d -\frac{b}{3}+ \epsilon^\prime &  -\frac{b}{3} & c +\frac{2b}{3}
    \end{pmatrix}
\end{equation}

where $\epsilon ^\prime$=$x_\epsilon^{\prime}\frac{u^{\prime} v_\rho^2}{\Lambda^3}$, parameterizes the correction to the TBM neutrino mixing. Applying similar condition on $c$ and $d$ as in case I, we obtain

\begin{equation}
\label{eq:11}
    M^{(II)}_\nu= m_0
    \begin{pmatrix}
    \frac{2b}{3} & d -\frac{b}{3} & d -\frac{b}{3}+ \epsilon^\prime \\
    d- \frac{b}{3}  & d +\frac{2b}{3}+ \epsilon^\prime & -\frac{b}{3}\\
    d -\frac{b}{3}+ \epsilon^\prime &  -\frac{b}{3} & d +\frac{2b}{3}
    \end{pmatrix}
\end{equation}

In section 3, we give the detailed phenomenological analysis for both the cases and discuss the effect of perturbations ($\epsilon$ and $\epsilon^\prime$) on various neutrino oscillation parameters. Further, we present a numerical study of neutrinoless double-beta decay considering the allowed parameter space of the model.

\section{Numerical Analysis and results} 

In the previous section, we have shown how Altarelli-Feruglio A$_4$ model could be modified by adding extra three singlet flavons and taking into consideration higher dimension terms. In this section, we perform a numerical analysis to study the capability of the perturbation parameters $\epsilon$ and $\epsilon^\prime$ to produce the deviation of neutrino mixing from exact TBM. For each case, we will discuss the results for both normal as well as inverted hierarchies. Throughout the numerical analysis, we have taken the value of $m_0$ to be in the range [0.016 - 0.032] eV.

\noindent The neutrino mass matrix $M^{(I)}_\nu$ and $M^{(II)}_\nu$ can be diagonalized by the PMNS matrix $U$ as
\begin{equation}
    \label{eq:12}
    U^\dagger M^{(i)}_\nu U^* = \textrm{diag(}m_1, m_2, m_3 \textrm{)}
\end{equation}
with $i$ = \{I, II\}. We can numerically calculate $U$ using the relation $U^\dagger h U = \textrm{diag(}m_1^2, m_2^2, m_3^2 \textrm{)}$, where $h = M^{(i)}_\nu M^{(i)\dagger}_\nu$. The neutrino oscillation parameters $\theta_{12}$, $\theta_{13}$, $\theta_{23}$ and $\delta_{CP}$ can be obtained from $U$ as
\begin{equation}
    \label{eq:13}
    s_{12}^2 = \frac{\lvert U_{12}\rvert ^2}{1 - \lvert U_{13}\rvert ^2}, ~~~~~~ s_{13}^2 = \lvert U_{13}\rvert ^2, ~~~~~~ s_{23}^2 = \frac{\lvert U_{23}\rvert ^2}{1 - \lvert U_{13}\rvert ^2},
\end{equation}

and $\delta$ may be given by
\begin{equation}
    \label{eq:14}
    \delta = \textrm{sin}^{-1}\left(\frac{8 \, \textrm{Im(}h_{12}h_{23}h_{31}\textrm{)}}{P}\right)
\end{equation}
with 
\begin{equation}
    \label{eq:15}
     P = (m_2^2-m_1^2)(m_3^2-m_2^2)(m_3^2-m_1^2)\sin 2\theta_{12} \sin 2\theta_{23} \sin 2\theta_{13} \cos \theta_{13}
\end{equation}

For the comparison of theoretical neutrino mixing parameters with the latest experimental data \cite{Esteban:2020cvm}, the modified $A_4$ model is fitted to the experimental data by minimizing the following $\chi^2$ function:

\begin{equation}
	\label{eq:16}
	\chi^2 = \sum_{i}\left(\frac{\lambda_i^{model} - \lambda_i^{expt}}{\Delta \lambda_i}\right)^2.
\end{equation}
where $\lambda_i^{model}$ is the $i^{th}$ observable predicted by the model, $\lambda_i^{expt}$ stands for the $i^{th}$ experimental best-fit value and $\Delta \lambda_i$ is the $1\sigma$ range of the $i^{th}$ observable.

\begin{figure}[t]
     \centering
     \begin{subfigure}[b]{0.46\textwidth}
         \centering
         \includegraphics[width=\textwidth]{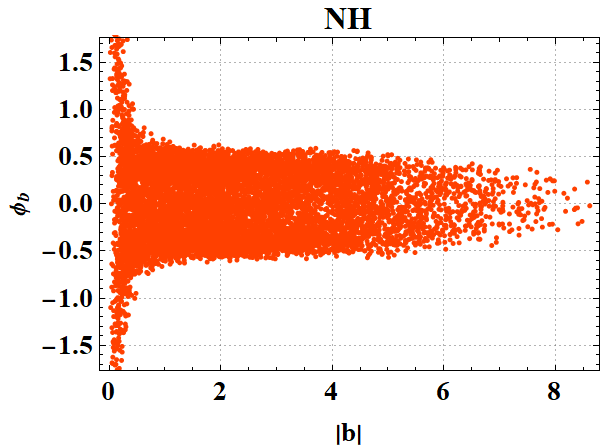}
     \end{subfigure}
     \hfill
     \begin{subfigure}[b]{0.46\textwidth}
         \centering
         \includegraphics[width=\textwidth]{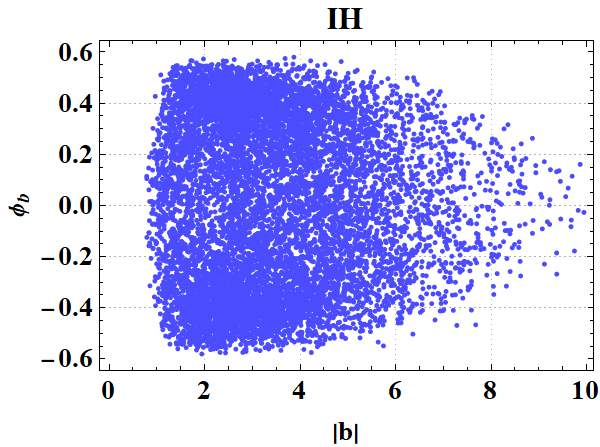}
     \end{subfigure}

     \vspace{1em}
     \begin{subfigure}[b]{0.46\textwidth}
         \centering
         \includegraphics[width=\textwidth]{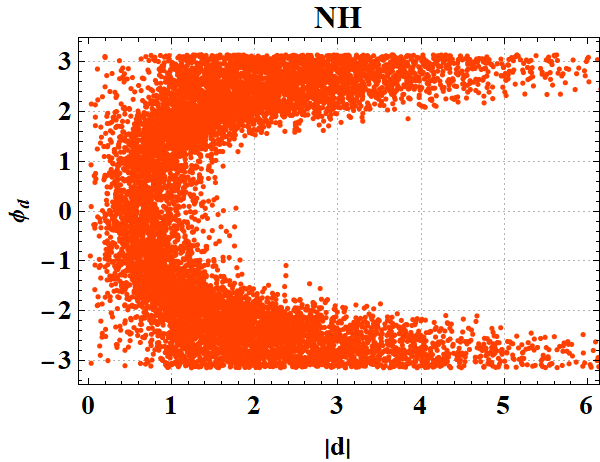}
     \end{subfigure}
     \hfill
     \begin{subfigure}[b]{0.46\textwidth}
         \centering
         \includegraphics[width=\textwidth]{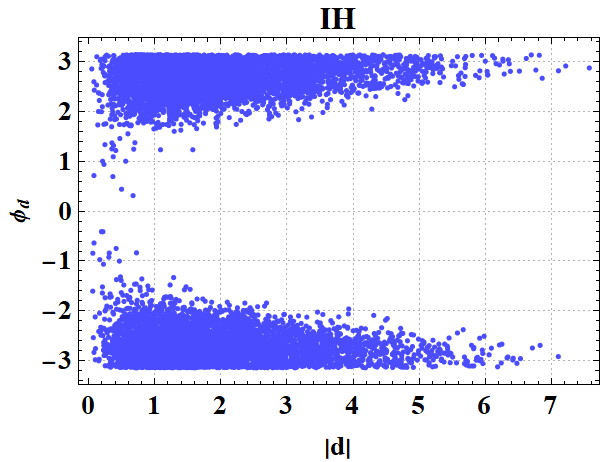}
     \end{subfigure}
    \caption{Case I - Correlation among the model parameters $\lvert b\rvert$, $\phi_b$ \& $\lvert d\rvert$, $\phi_d$ in both NH (left column) as well as IH (right column).}
    \label{fig:1}
\end{figure}

\begin{figure}[!ht]
     \centering
     \begin{subfigure}[b]{0.46\textwidth}
         \centering
         \includegraphics[width=\textwidth]{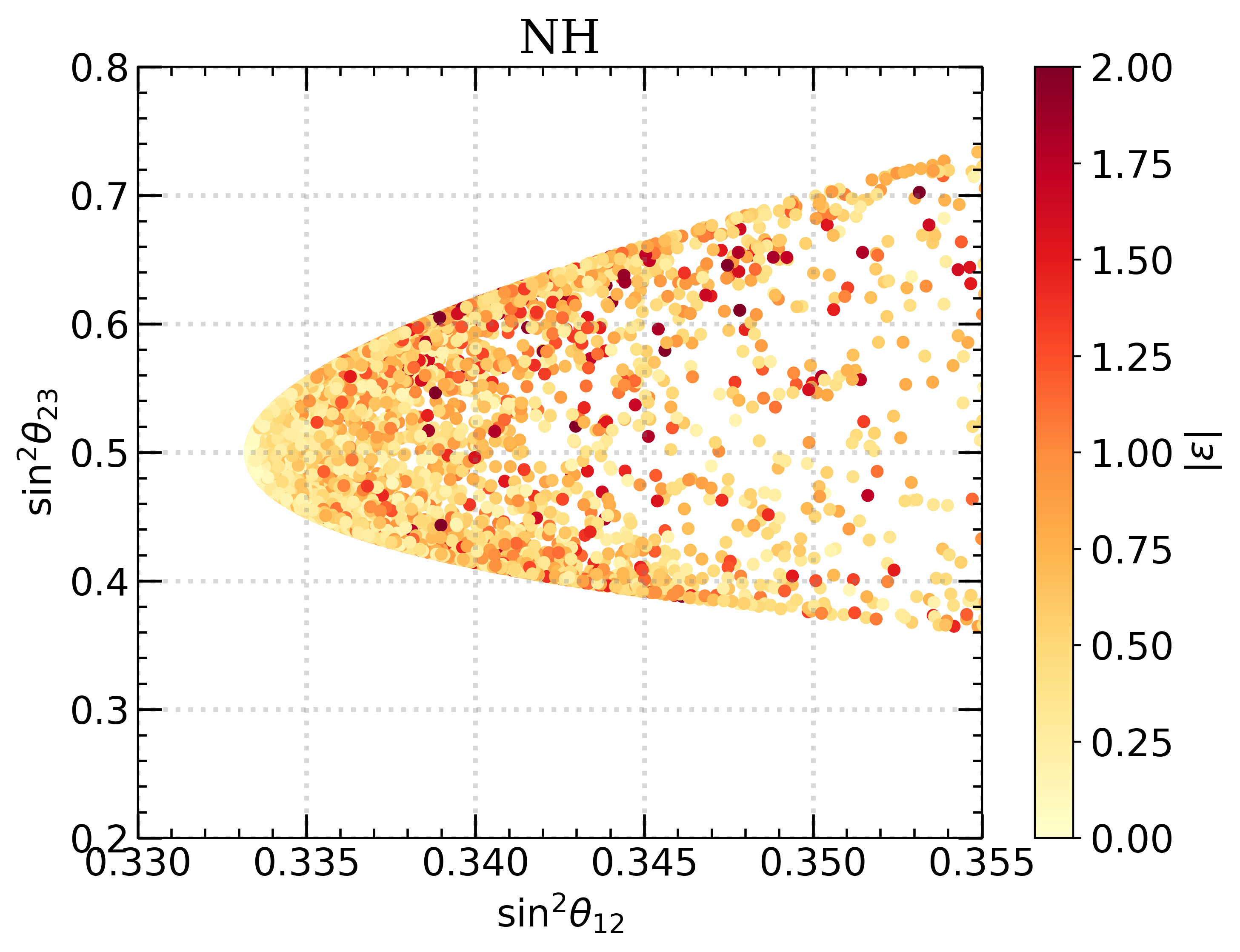}
     \end{subfigure}
     \hfill
     \begin{subfigure}[b]{0.46\textwidth}
         \centering
         \includegraphics[width=\textwidth]{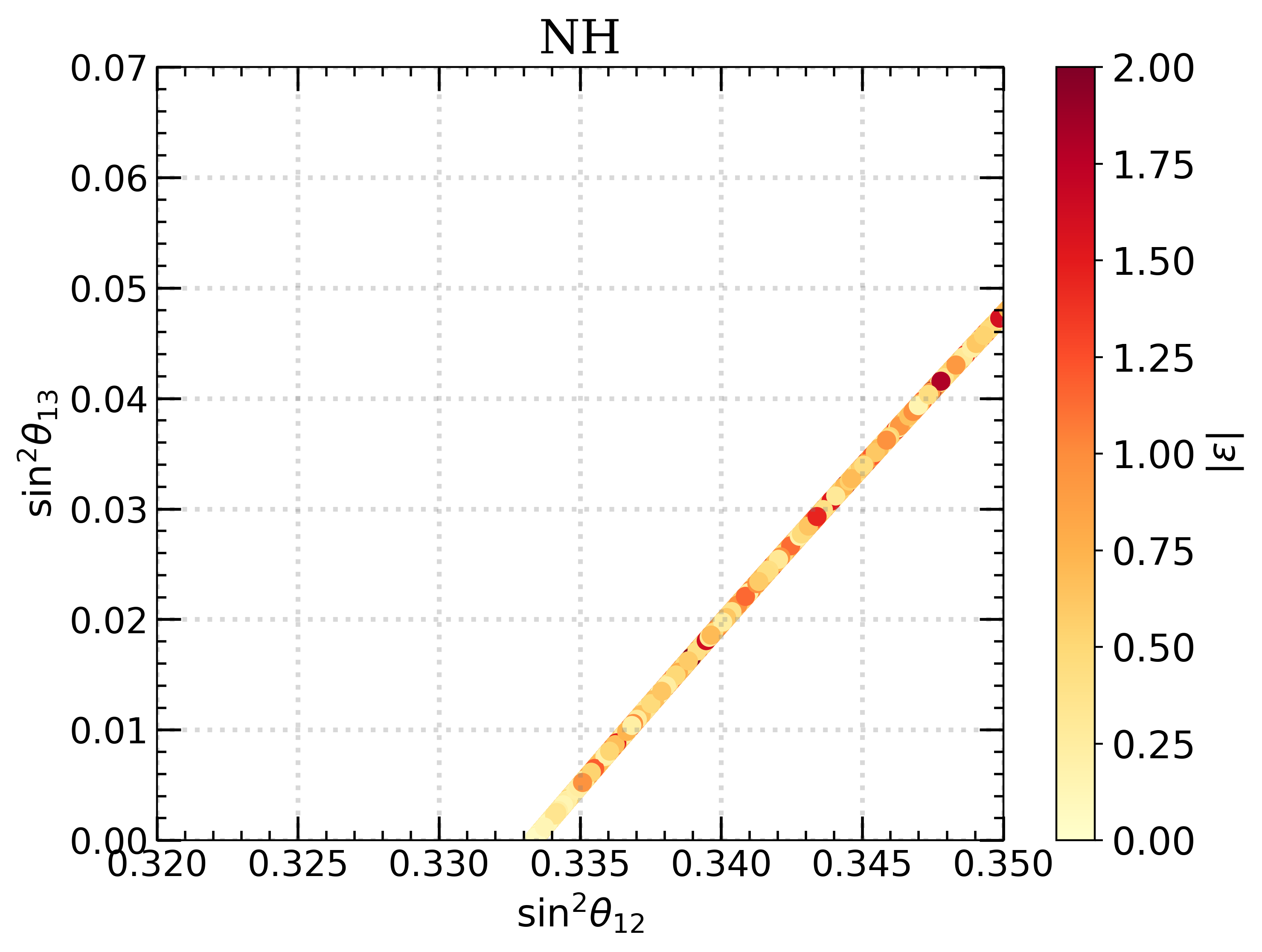}
     \end{subfigure}

     \vspace{1em}
     \begin{subfigure}[b]{0.46\textwidth}
         \centering
         \includegraphics[width=\textwidth]{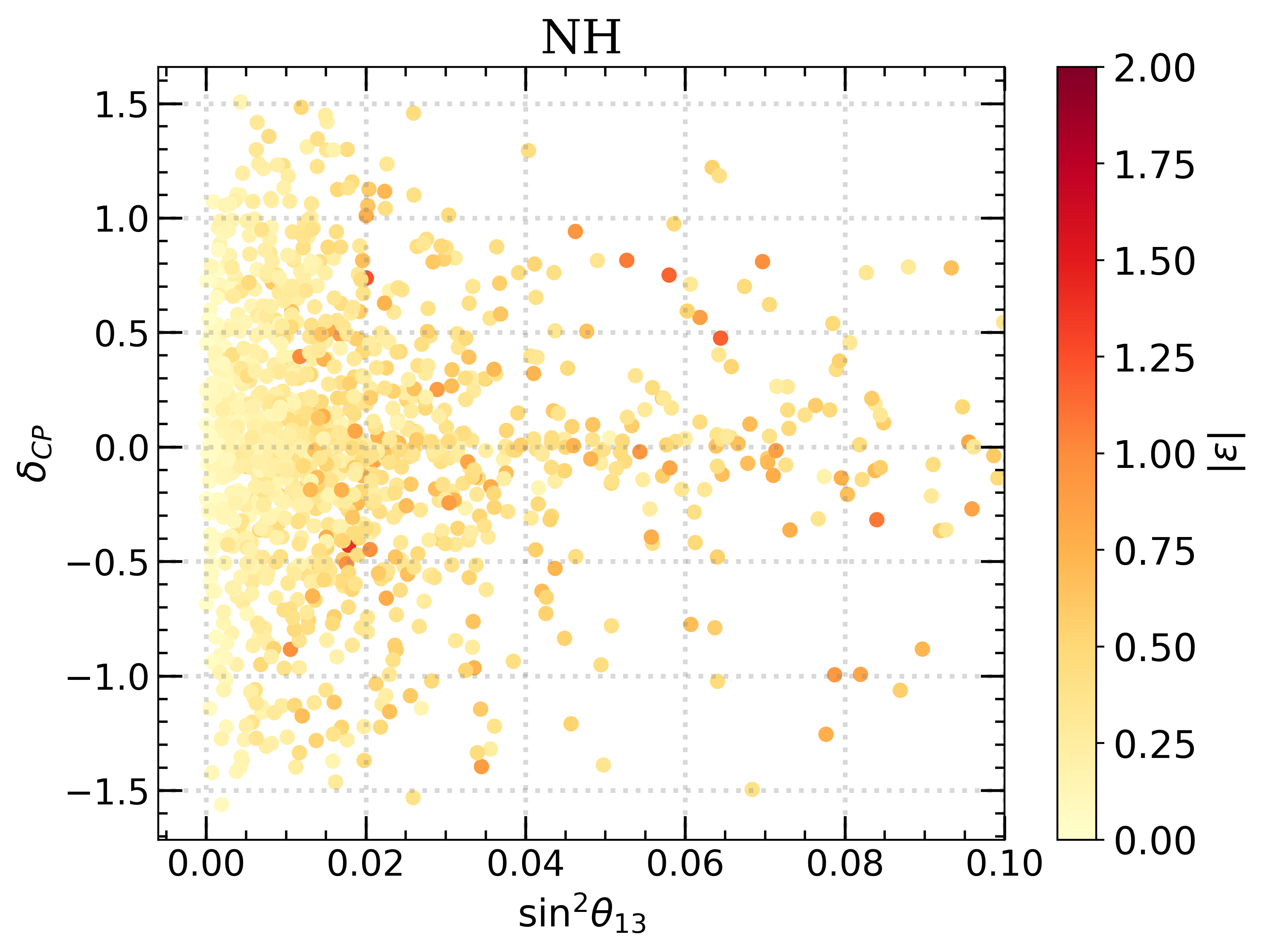}
     \end{subfigure}
     \hfill
     \begin{subfigure}[b]{0.46\textwidth}
         \centering
         \includegraphics[width=\textwidth]{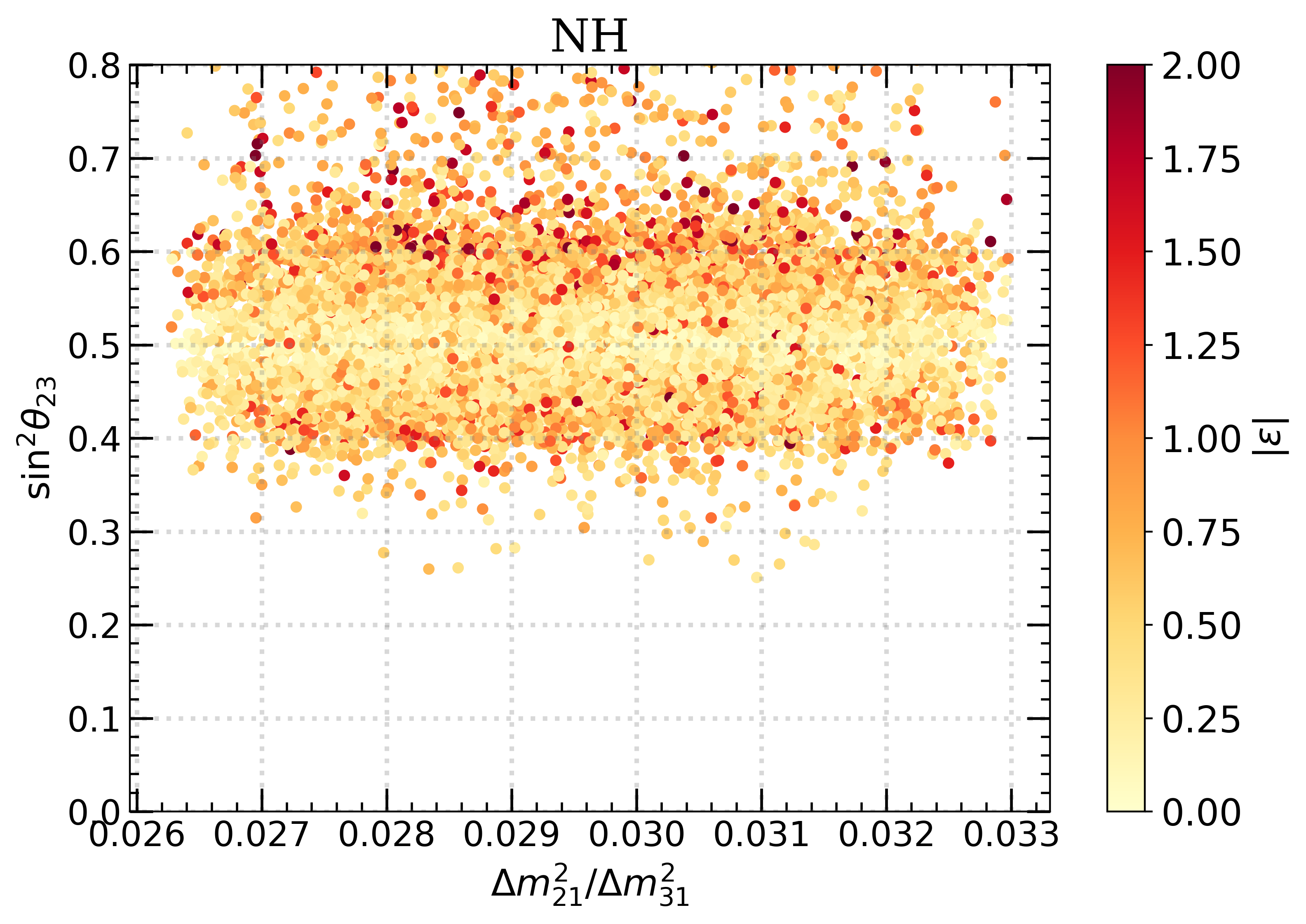}
     \end{subfigure}
    \caption{Case I - Variation of the mixing angles, mass-squared differences and Dirac CP phase with the correction parameter $\epsilon$ in NH case. The color-bar represents the different values of the correction parameter $\epsilon$.}
    \label{fig:2}
\end{figure}

First, we shall discuss case I, for perturbation parameter $\epsilon$.
In Fig. \ref{fig:1}, we have shown the parameter space of the model for Case I, which is constrained using the 3$\sigma$ bound on neutrino oscillation data (Table \ref{tab:1}). For both normal and inverted hierarchies, one can see that there is a high correlation between different parameters of the model.

Fig. \ref{fig:2} shows the prediction of the various neutrino oscillation parameters for NH in case I. The calculated best fit values of $\sin^2 \theta_{12}$, $\sin^2 \theta_{13}$ and $\sin^2 \theta_{23}$ are (0.341, 0.023, 0.62) which are within the 3$\sigma$ range of experimental values. Other parameters such as $\Delta m_{21}^2$, $\Delta m_{31}^2$ and $\delta_{CP}$ have their best-fit values, corresponding to $\chi^2$-minimum, at ($7.422 \times 10^{-5} eV^2$, $2.556 \times 10^{-3} eV^2$, $-0.367 \pi$) respectively, which perfectly agreed with the latest observed neutrino oscillation experimental data. Similar variation of $\sin^2 \theta_{13}$ and  $\sin^2 \theta_{23}$ with increase in $\lvert \epsilon \rvert$ is observed for IH.

\begin{figure}[t]
     \centering
     \begin{subfigure}[b]{0.46\textwidth}
         \centering
         \includegraphics[width=\textwidth]{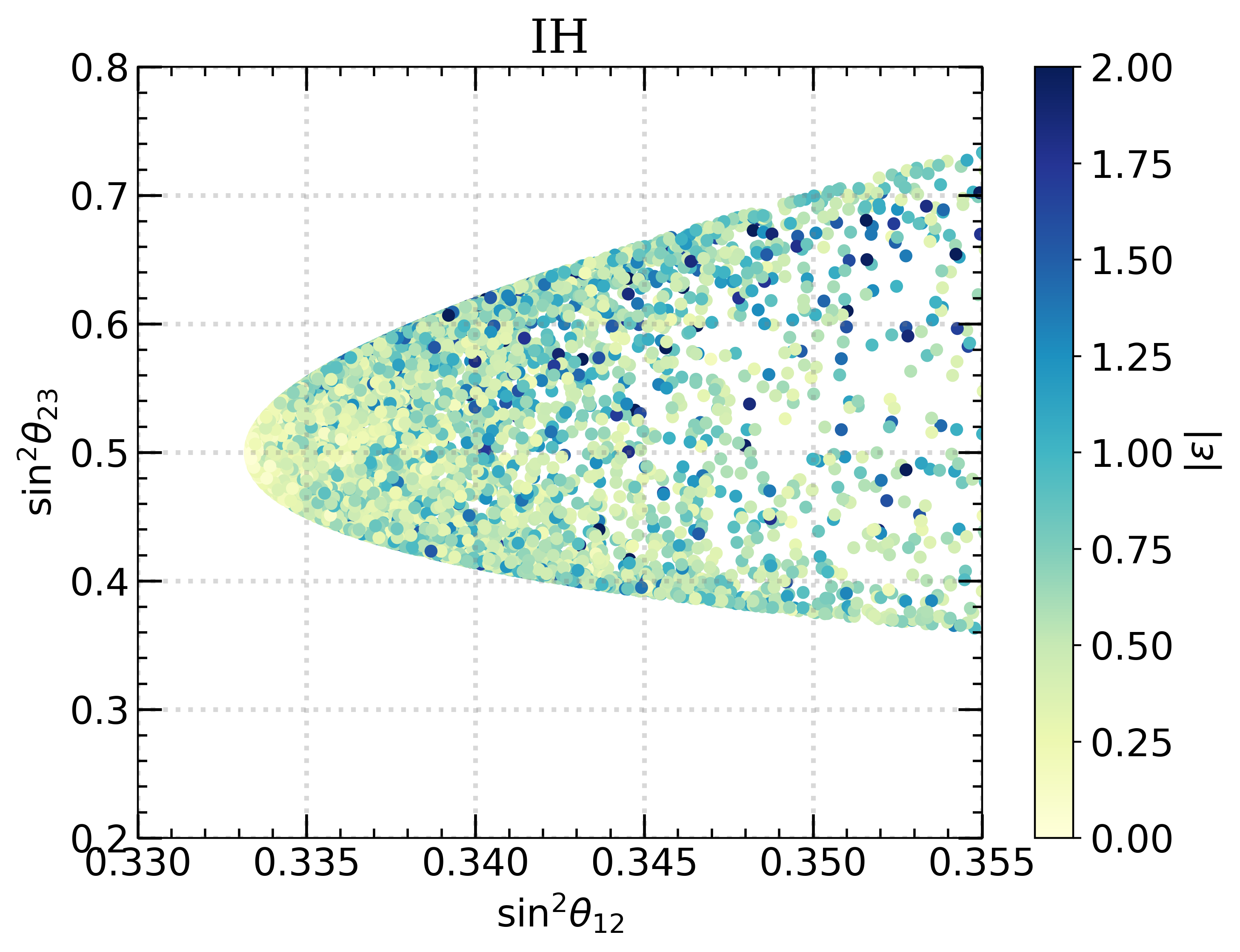}
     \end{subfigure}
     \hfill
     \begin{subfigure}[b]{0.46\textwidth}
         \centering
         \includegraphics[width=\textwidth]{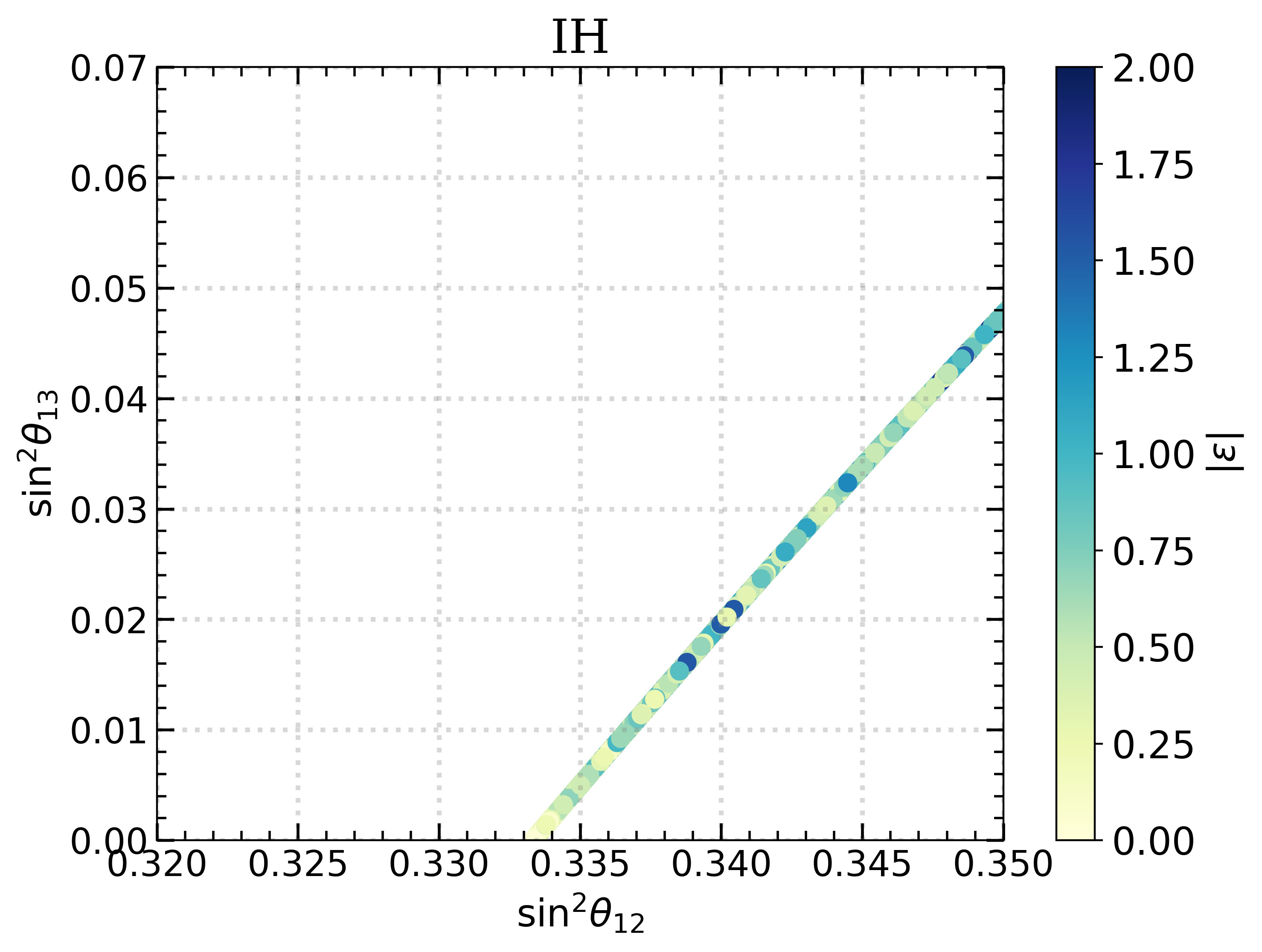}
     \end{subfigure}

     \vspace{1em}
     \begin{subfigure}[b]{0.46\textwidth}
         \centering
         \includegraphics[width=\textwidth]{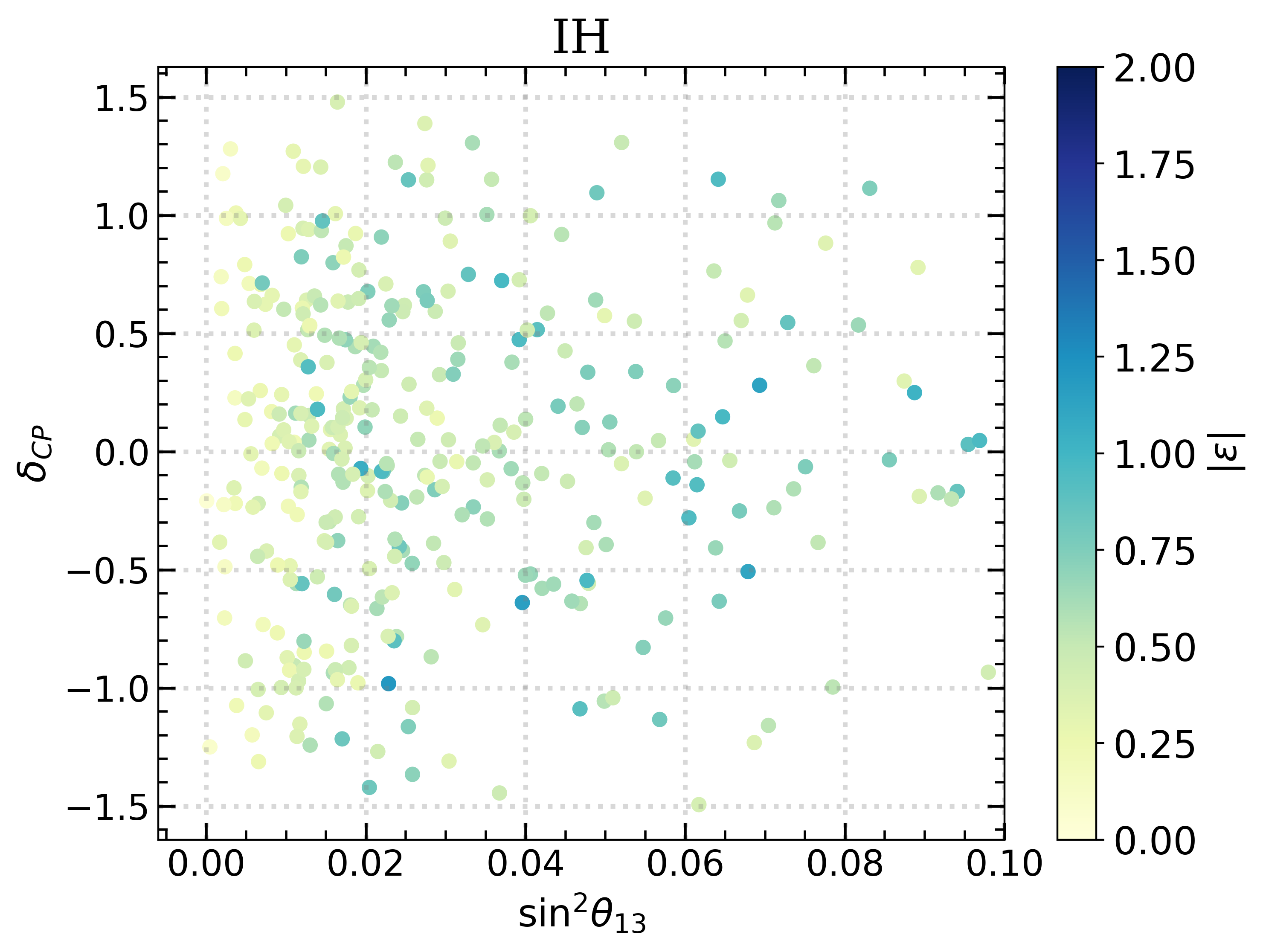}
     \end{subfigure}
     \hfill
     \begin{subfigure}[b]{0.46\textwidth}
         \centering
         \includegraphics[width=\textwidth]{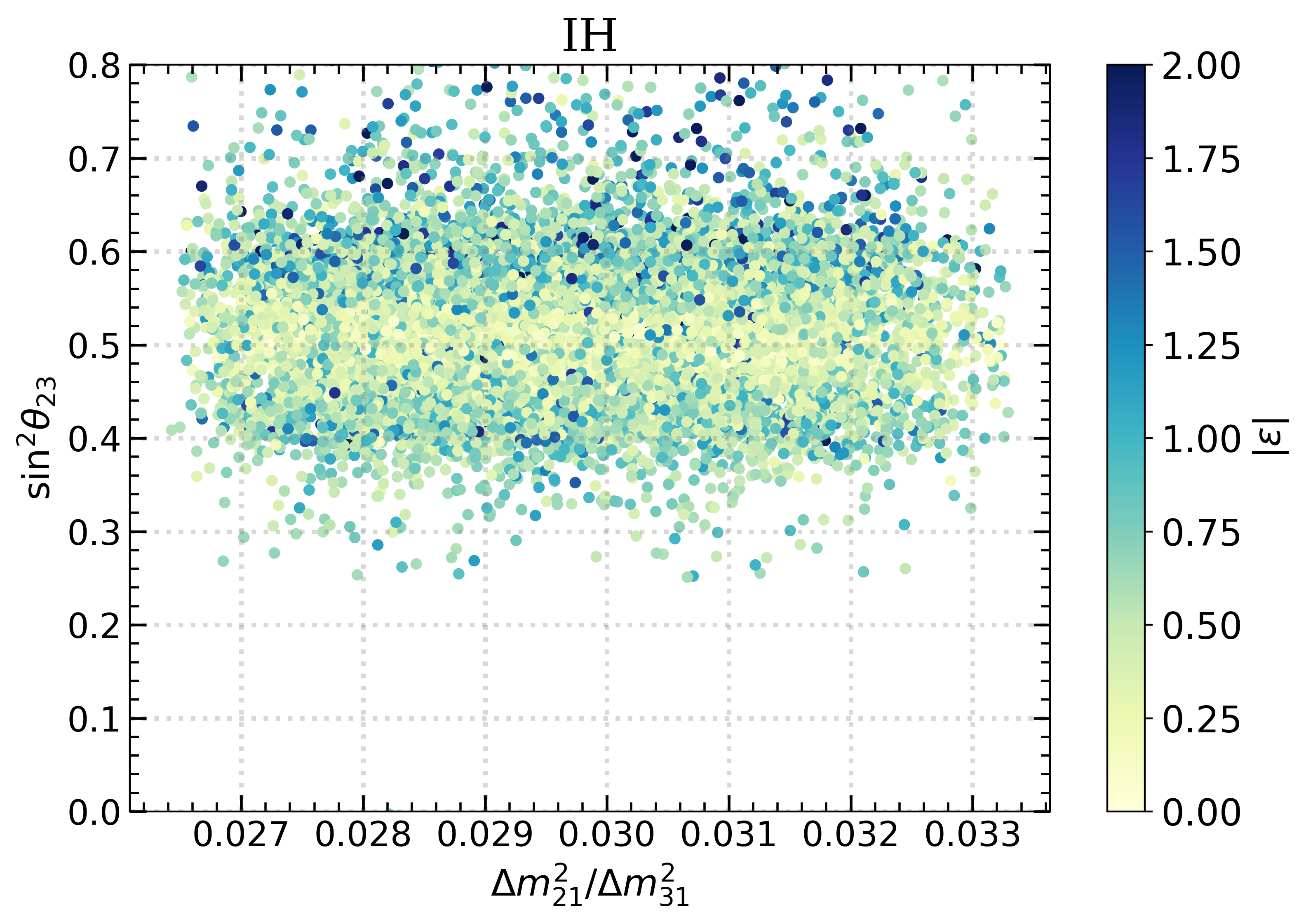}
     \end{subfigure}
    \caption{Case I - Variation of the mixing angles, mass-squared differences and Dirac CP phase with the correction parameter $\epsilon$ in IH case. The color-bar represents the different values of the correction parameter $\epsilon$.}
    \label{fig:3}
\end{figure}

Fig. \ref{fig:3} gives the neutrino oscillation parameters predicted by the model for IH. (0.341, 0.024, 0.61) are the best fit values of $\sin^2 \theta_{12}$, $\sin^2 \theta_{13}$ and $\sin^2 \theta_{23}$, which are all within the 3 $\sigma$ range of experimental values. Also, $\Delta m_{21}^2$, $\Delta m_{31}^2$ and $\delta_{CP}$ have their best-fit values, corresponding to $\chi^2$-minimum, at ($7.465 \times 10^{-5} eV^2$, $2.492 \times 10^{-3} eV^2$, $ 0.39 \pi$). Besides, we determined that, as the correction parameter $\lvert \epsilon \rvert$ increases, the value of $\sin^2 \theta_{13}$ and  $\sin^2 \theta_{23}$ moves away from 0 and $\frac{1}{2}$, respectively.

Thus, the model defined in Case I, clearly shows that the deviation from exact tri-bimaximal mixing, however, with the change in $\lvert \epsilon \rvert$ there is no observable preference for the octant of $\theta_{23}$.

\begin{figure}[t]
     \centering
     \begin{subfigure}[b]{0.46\textwidth}
         \centering
         \includegraphics[width=\textwidth]{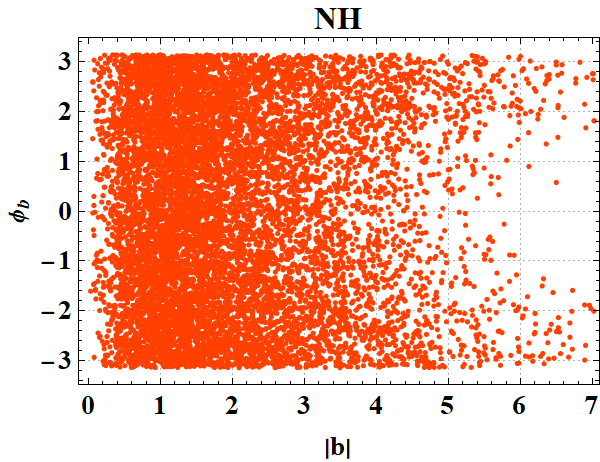}
     \end{subfigure}
     \hfill
     \begin{subfigure}[b]{0.46\textwidth}
         \centering
         \includegraphics[width=\textwidth]{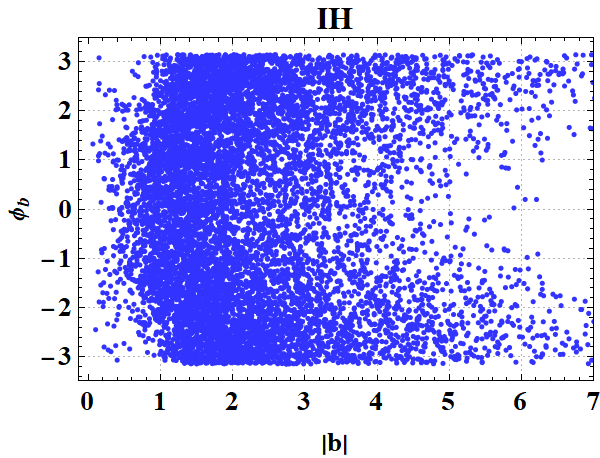}
     \end{subfigure}

     \vspace{1em}
     \begin{subfigure}[b]{0.46\textwidth}
         \centering
         \includegraphics[width=\textwidth]{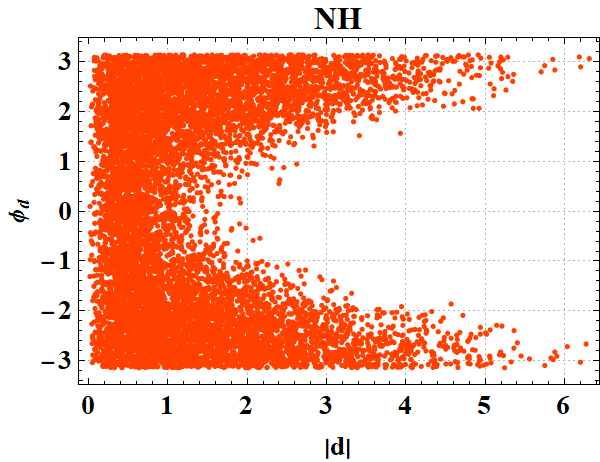}
     \end{subfigure}
     \hfill
     \begin{subfigure}[b]{0.46\textwidth}
         \centering
         \includegraphics[width=\textwidth]{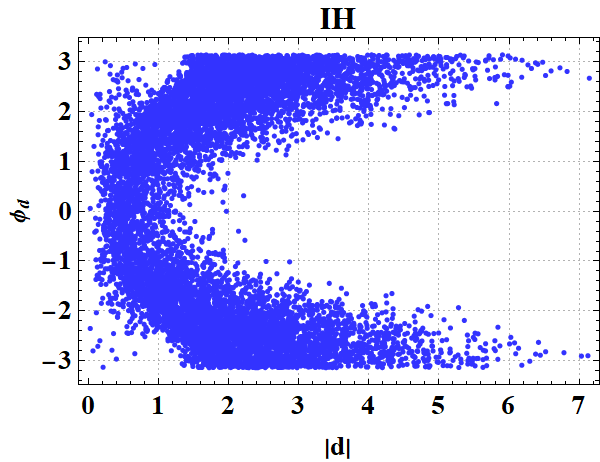}
     \end{subfigure}
    \caption{Case II - Correlation among the model parameters $\lvert b\rvert$, $\phi_b$ \& $\lvert d\rvert$, $\phi_d$ in both NH (left column) as well as IH (right column).}
    \label{fig:4}
\end{figure}

\begin{figure}[!ht]
     \centering
     \begin{subfigure}[b]{0.4\textwidth}
         \centering
         \includegraphics[width=\textwidth]{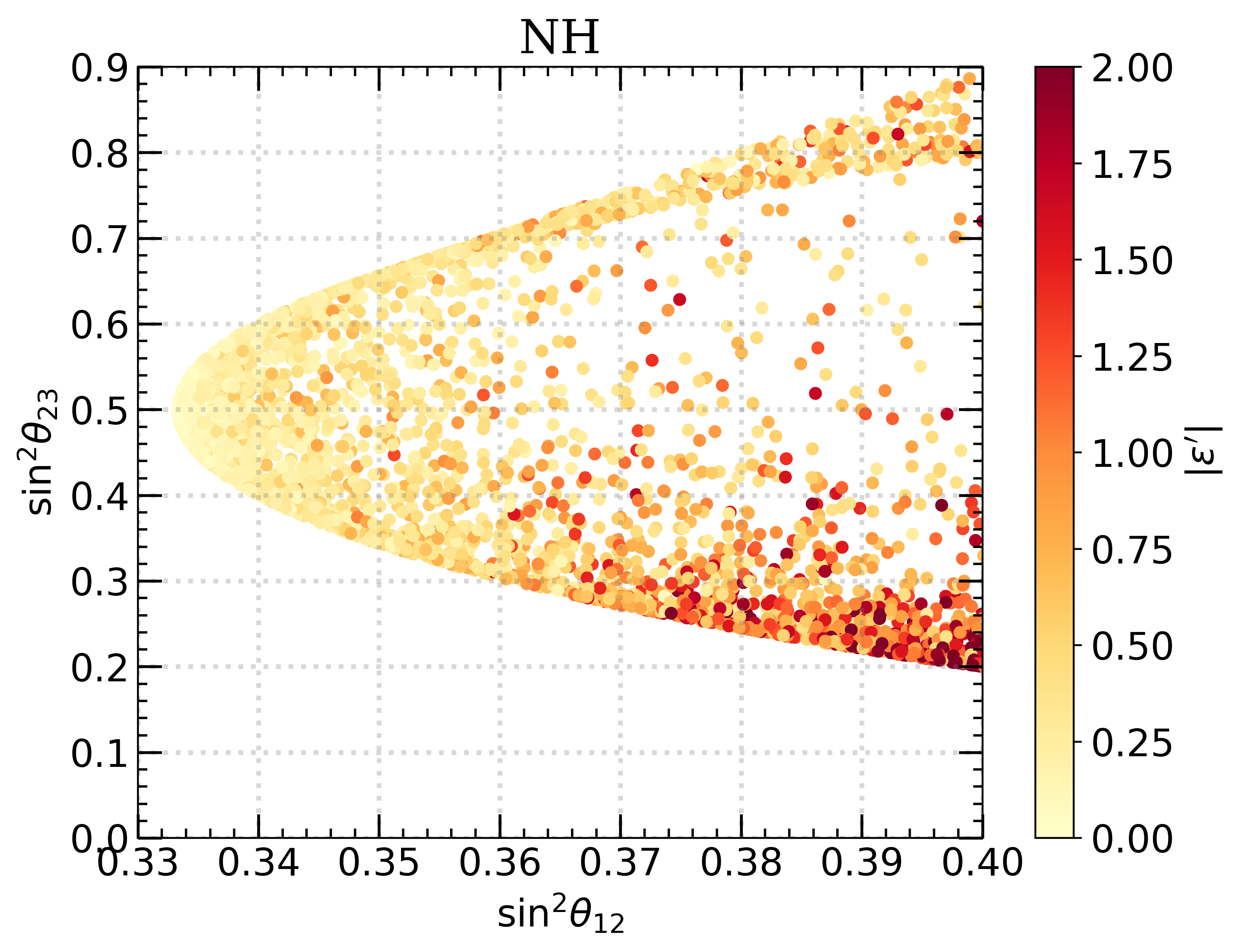}
     \end{subfigure}
     \hfill
     \begin{subfigure}[b]{0.4\textwidth}
         \centering
         \includegraphics[width=\textwidth]{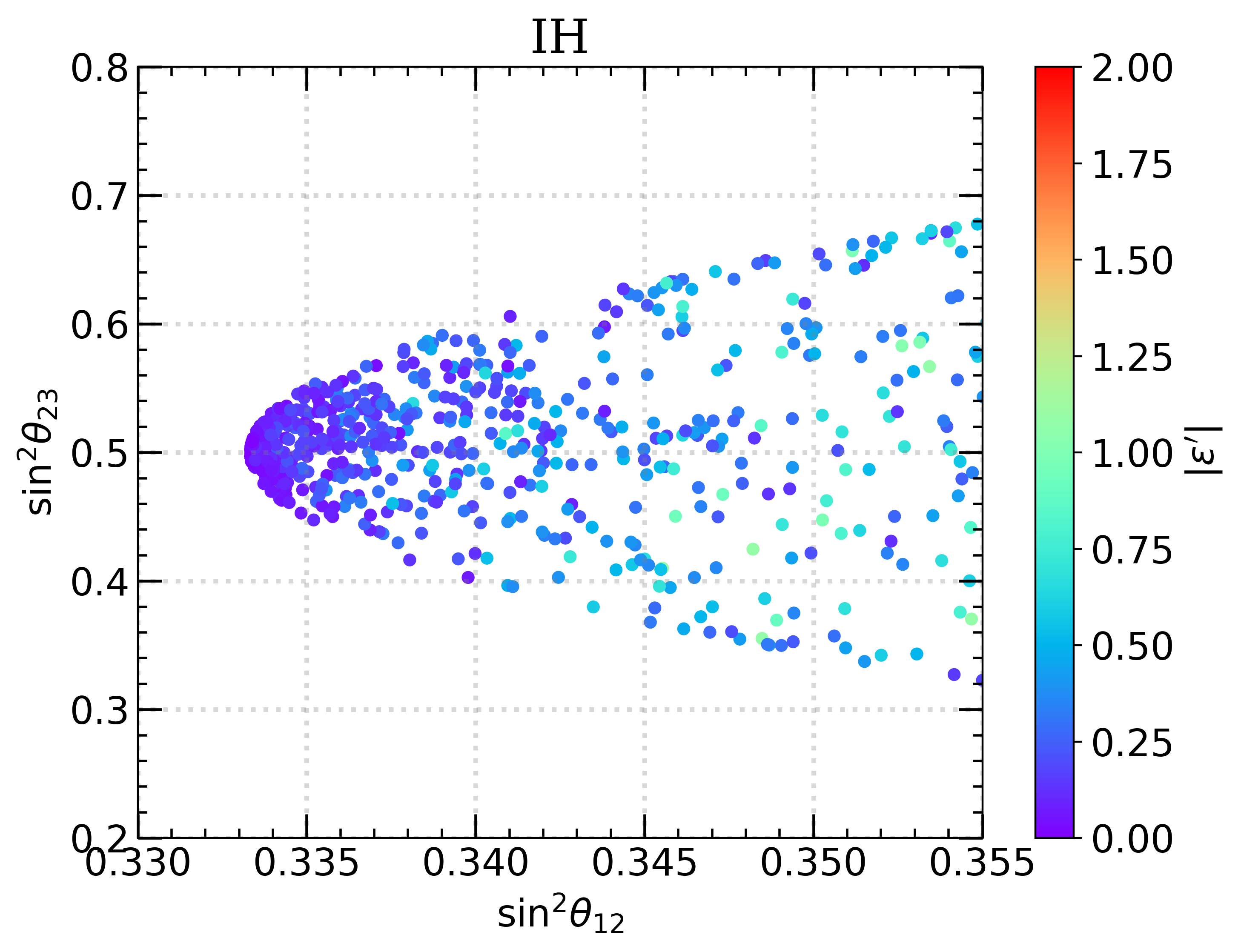}
     \end{subfigure}
     \begin{subfigure}[b]{0.4\textwidth}
         \centering
         \includegraphics[width=\textwidth]{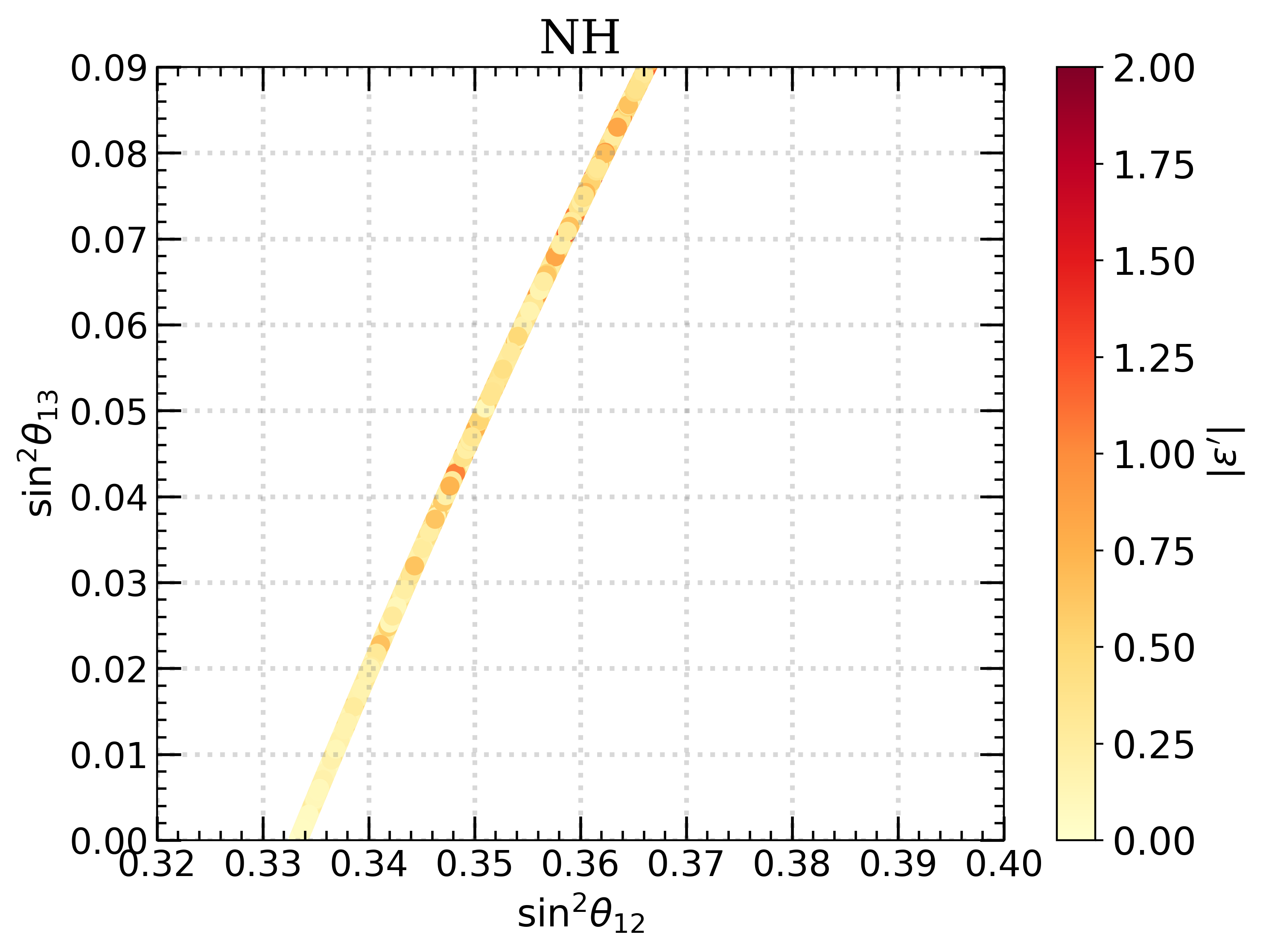}
     \end{subfigure}
     \hfill
     \begin{subfigure}[b]{0.4\textwidth}
         \centering
         \includegraphics[width=\textwidth]{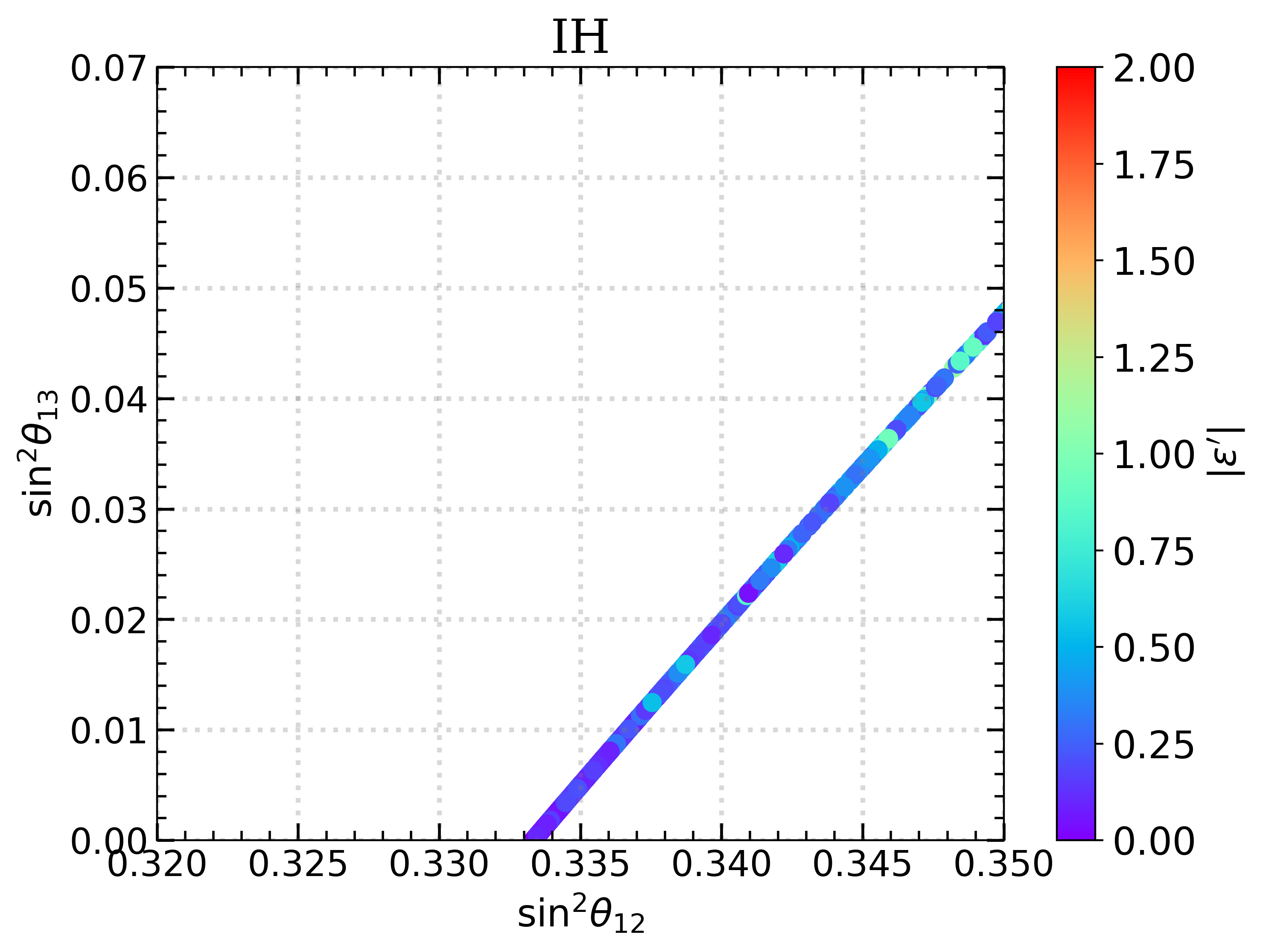}
     \end{subfigure}
     
          \begin{subfigure}[b]{0.4\textwidth}
         \centering
         \includegraphics[width=\textwidth]{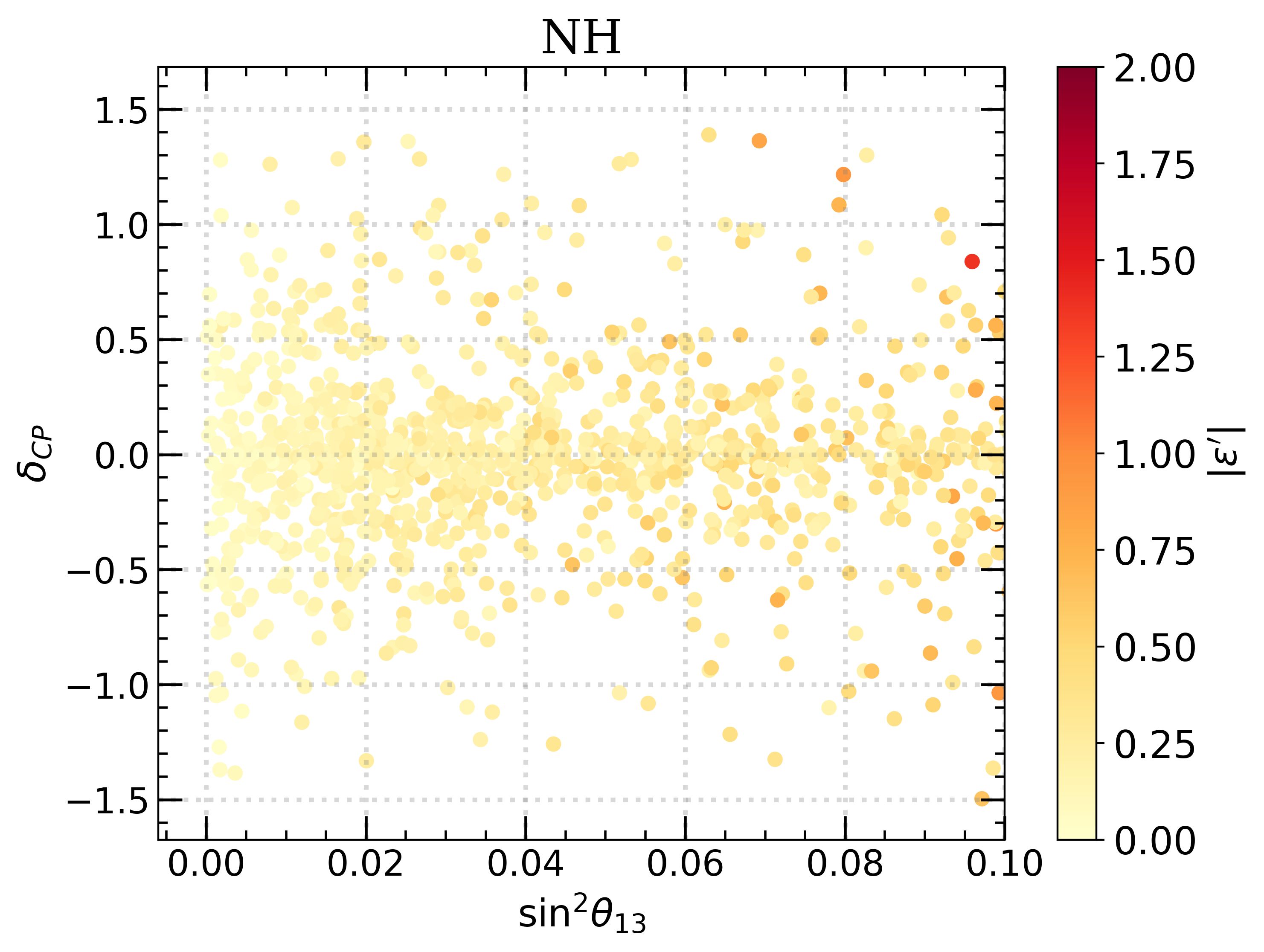}
     \end{subfigure}
     \hfill
     \begin{subfigure}[b]{0.4\textwidth}
         \centering
         \includegraphics[width=\textwidth]{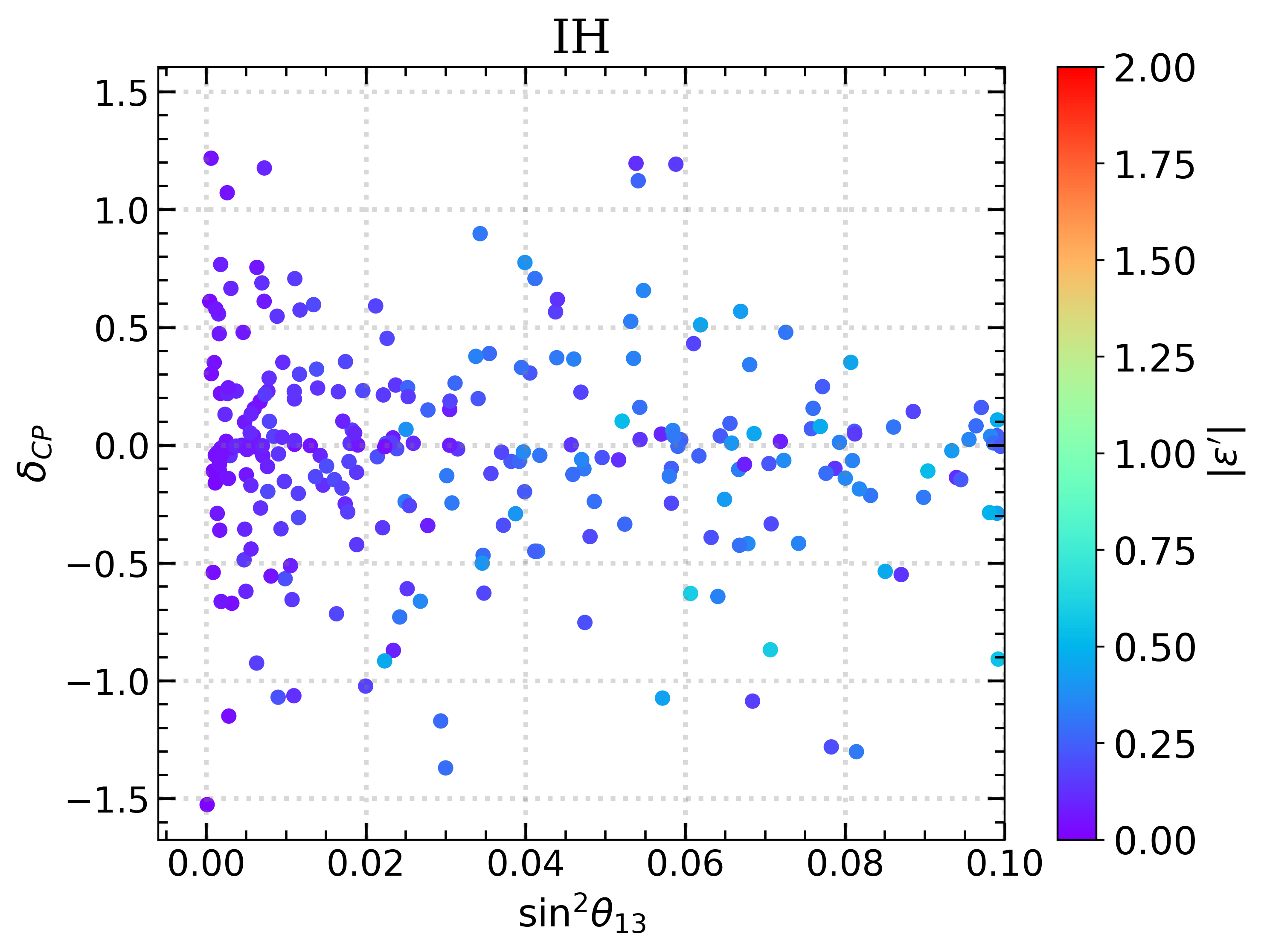}
     \end{subfigure}
    
     \begin{subfigure}[b]{0.4\textwidth}
         \centering
         \includegraphics[width=\textwidth]{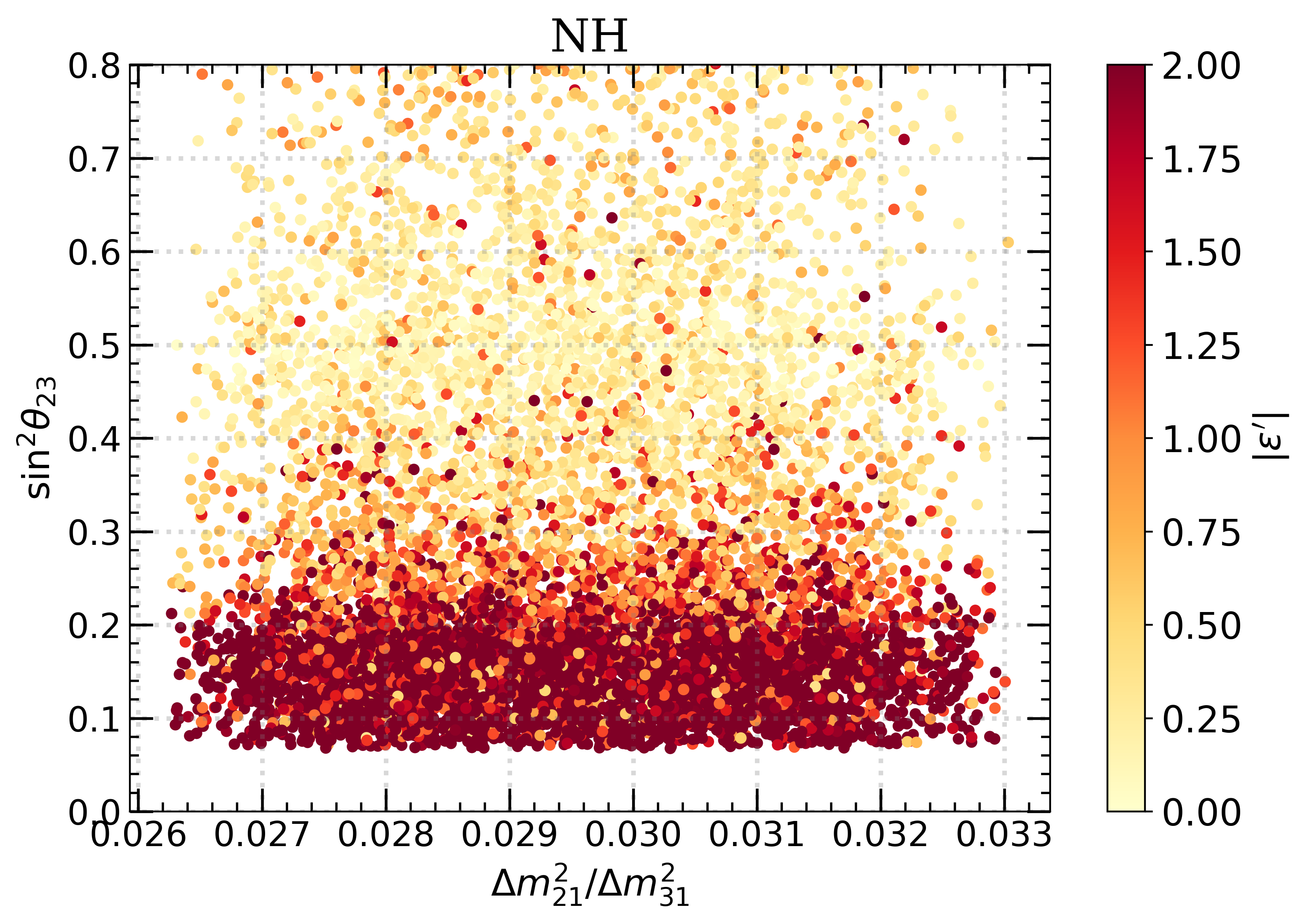}
     \end{subfigure}
     \hfill
     \begin{subfigure}[b]{0.4\textwidth}
         \centering
         \includegraphics[width=\textwidth]{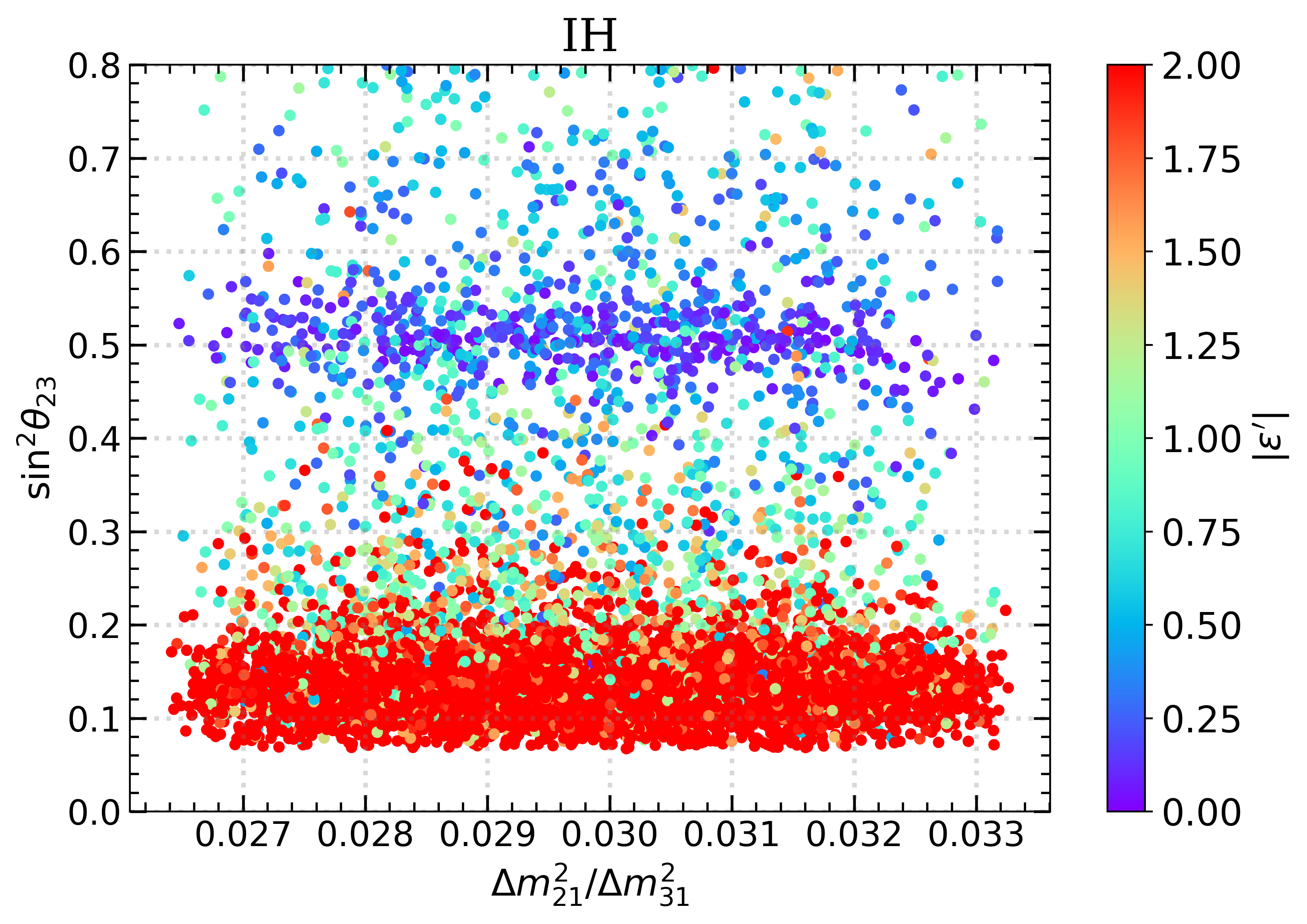}
     \end{subfigure}
    \caption{Case II - Variation of the mixing angles, mass-squared differences and Dirac CP phase with the correction parameter $\epsilon^\prime$ in NH (left) and IH(right) case. The color-bar represents the different values of the correction parameter $\epsilon^\prime$.}
    \label{fig:5}
\end{figure}

Now, we discuss for the case II for the perturbation parameter $\epsilon^\prime$. In Figs. \ref{fig:4} and \ref{fig:5}, we show the results with second type of correction, described as Case II in section \ref{sec:2}. The allowed ranges for the model parameters for both NH as well as IH are presented in Fig. \ref{fig:4}. Also, the different correlation plots among the various neutrino oscillation parameters and their variations with $\lvert \epsilon^\prime\rvert$ are shown in fig. \ref{fig:5}. It is clear that the neutrino mixing deviates from exact TBM mixing as $\lvert \epsilon^\prime\rvert$ increases from 0 in both NH and IH cases. The prediction of the atmospheric mixing angle $\theta_{23}$ in NH case shows slight preference towards lower octant for larger values of $\epsilon^\prime$. Thus, the modification of Altarelli-Feruglio with additional term, $x_\epsilon^{\prime}\xi^{\prime} {(l l)}^{\prime\prime}\frac{\rho\rho}{\Lambda^2}$, allows us to deviate from TBM mixing. The best-fit values for the predictions of the various oscillation parameters is shown in Table \ref{tab:3}.

\begin{table}[!ht]
    \centering
    \begin{tabular}{c c c}
    \hline
       \textrm{Parameter}  &  \textrm{NH} & \textrm{IH}\\
     \hline
     
     $\sin^2\theta_{12}$ & 0.3407 & 0.341\\
     
     $\sin^2\theta_{13}$ & 0.0217 & 0.0226\\
     
     $\sin^2\theta_{23}$ & 0.576 & 0.547\\
     
     $\delta_{CP}$ & 0.061 $\pi$ & 0.145 $\pi$ \\
     
     $\Delta m_{21}^2$ & $7.498 \times 10^{-5}$ & $7.492 \times 10^{-5}$\\
     
     $\Delta m_{31}^2$ & $2.517 \times 10^{-3}$ & $2.484 \times 10^{-5}$\\

    \hline
    \end{tabular}
    \caption{Case II- Best-fit values for different parameters predicted by the model }
    \label{tab:3}
\end{table}

\subsection{Neutrinoless double beta decay (NDBD):}
Additionally, we are unsure about the  nature of neutrinos - whether Majorana or Dirac. For neutrino to be Majorana in nature, the study of NDBD is very important. There are some ongoing NDBD experiments to determine Majorana nature of neutrino.
The effective mass that governs the process is provided by,
\begin{equation}
m_{\beta\beta}= U^2_{Li} m_i
\end{equation}
where $U_{Li}$ are the elements of the first row of the neutrino mixing matrix $U_{PMNS}$
(Eq.\ref{eq:1}) which is dependent on known parameters $\theta_{12}$, $\theta_{13}$ and the unknown Majorana phases $\alpha$ and $\beta$. $U_{PMNS}$ is the diagonalizing matrix of the light neutrino mass matrix $m_\nu$ so that,
\begin{equation}
m_\nu= U_{PMNS} M^{(diag)}_\nu U^T_{PMNS}
\end{equation}
where, $M^{(diag)}_\nu$ =diag($m_1$, $m_2$, $m_3$). The effective Majorana mass can be parameterized using the diagonalizing matrix elements and the mass eigen values as follows:
\begin{equation}
m_{\beta\beta}= m_1 c_{12}^2 c^2_{13}+ m_2 s^2_{12} c^2_{13} e^{2 i \alpha} + m_3 s^2_{13} e^{2i \beta}
\end{equation}

\begin{figure}[t]
     \centering
     \begin{subfigure}[b]{0.46\textwidth}
         \centering
         \includegraphics[width=\textwidth]{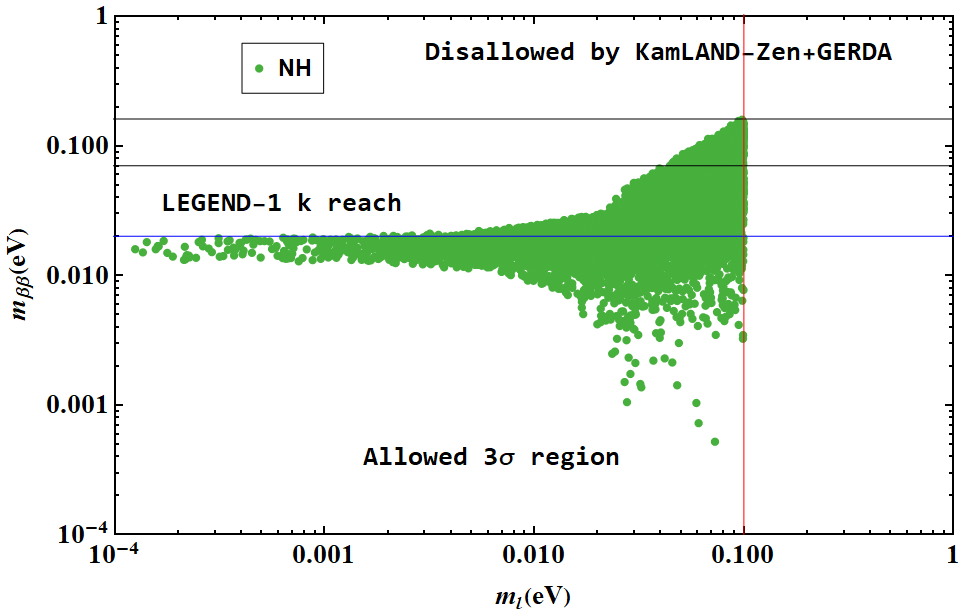}
     \end{subfigure}
     \hfill
     \begin{subfigure}[b]{0.46\textwidth}
         \centering
         \includegraphics[width=\textwidth]{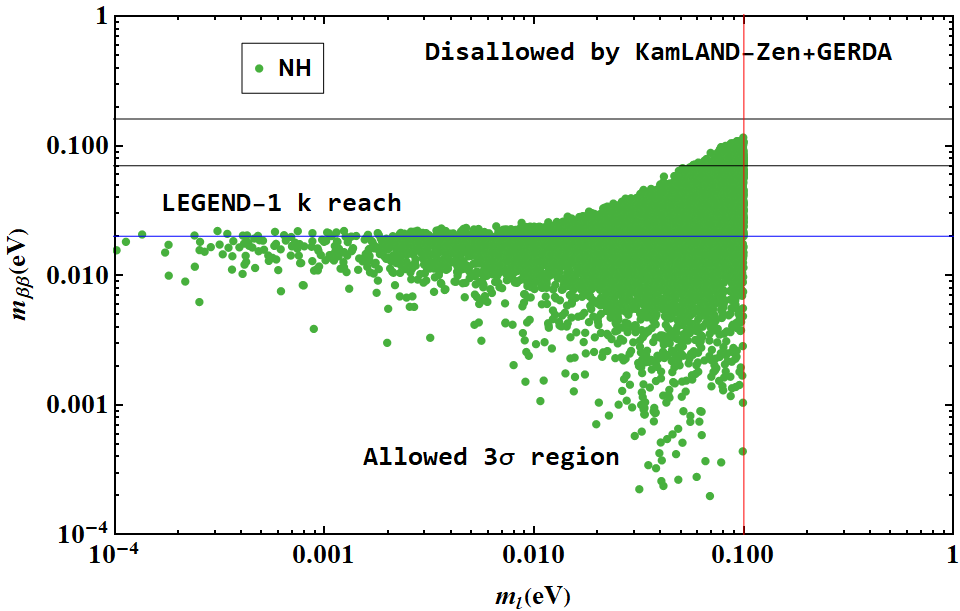}
     \end{subfigure}

     \vspace{1em}
     \begin{subfigure}[b]{0.46\textwidth}
         \centering
         \includegraphics[width=\textwidth]{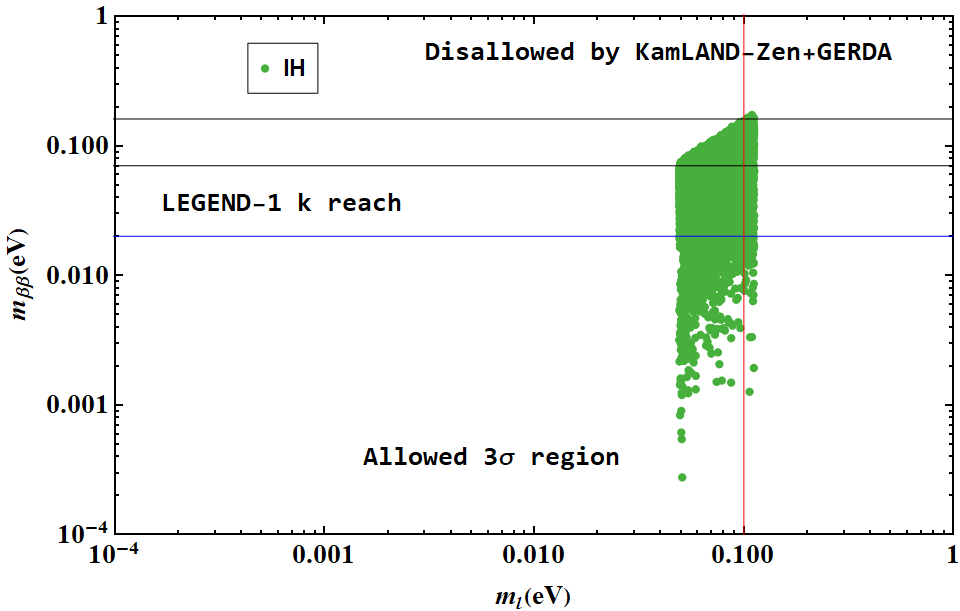}
     \end{subfigure}
     \hfill
     \begin{subfigure}[b]{0.46\textwidth}
         \centering
         \includegraphics[width=\textwidth]{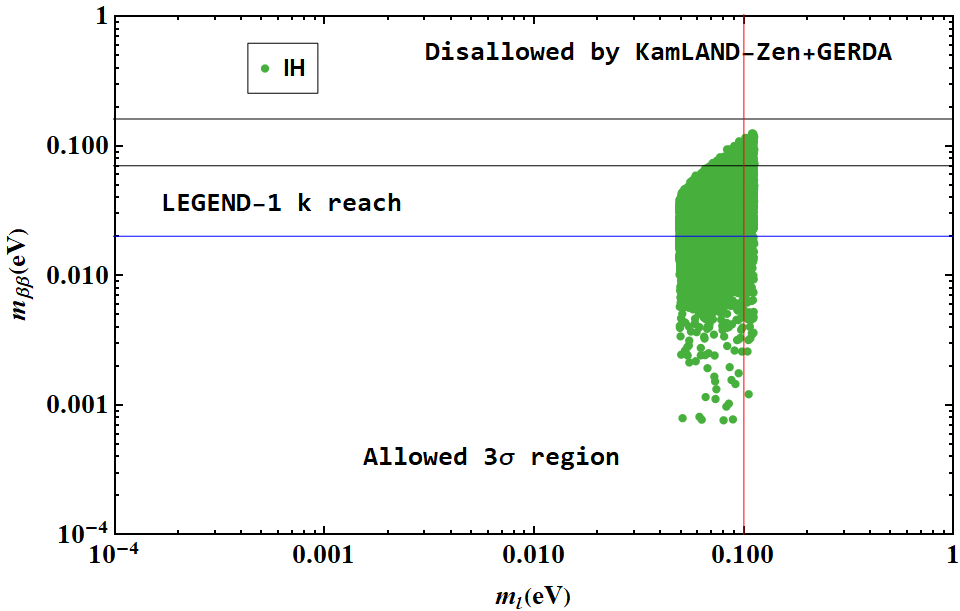}
     \end{subfigure}
    \caption{ Variation of effective Majorana neutrino mass with lighest neutrino mass in case I (left panel) and case II (right panel). The red solid line represents the cosmological bound on the sum of the absolute neutrino mass. The black solid line represents the KamLAND-Zen bound on the effective mass. }
    
    \label{fig:6}
\end{figure}

Using the constrained parameter space, we have evaluated the value of $m_{\beta\beta}$
for case I and case II in both NH as well as IH cases. The variation of $m_{\beta\beta}$
with lightest neutrino mass is shown in Figure \ref{fig:6} for both the neutrino mass hierarchies.
The sensitivity reach of NDBD experiments like KamLAND-Zen \cite{Gando:2020cxo, Biller:2021bqx}, GERDA \cite{DiMarco:2020mdk, GERDA:2017wlm, DAndrea:2020ezp},
LEGEND-1k \cite{abgrall2021legend} is also shown in figure 6. $m_{\beta\beta}$ is found to be well within the sensitivity reach of these NDBD experiments for both the cases, (I) and (II).

\section{Conclusion}

We have constructed a flavon-symmetric $A_4 \times Z_2\times Z_3$ model to realize the latest neutrino oscillation experimental data which depart from Tribimaximal neutrino mixing pattern. The models have been checked for accuracy by adding three extra  flavons $\xi^\prime$, $\xi^{\prime\prime}$, and $\rho$ transforming under representations of $A_4$ to extend the A-F model. And the cyclic $Z_2\times Z_3$ symmetric term has been incorporated to eliminate unwanted terms in the calculations. The calculated two perturbation parameters $\epsilon$ and $\epsilon^\prime$ clearly showed the deviation of neutrino mixing parameters from exact Tri-bimaximal neutrino mixing matrix. The resulting mass matrices give predictions for the neutrino oscillation parameters and their best-fit values are obtained using the $\chi^2$-analysis, which are consistent with the latest global neutrino oscillation experimental data. We found the magnitude of deviations from TBM is dominated by VEV of $\rho$. However, these quantities and corrections have little discriminative power, hence we supplement with observables related to neutrino mass. Therefore, we have also investigated NDBD in our model. The scatter plots of NDBD parameter ($m_{\beta\beta}$) and the lightest neutrino mass ($m_l$) parameter space are different in each model and allowed us to distinguish different models. The value of effective Majorana neutrino mass $|m_{\beta\beta}|$ is well within the sensitivity reach of the recent NDBD  experiments like KamLAND- Zen, GERDA and LEGEND-1k. The determination of NDBD, cosmological mass and leptonic CP-violation phase $\delta_{CP}$ which are consistent with the latest experimental data will discriminate the neutrino mass models.

\begin{acknowledgments}
Animesh Barman is grateful and acknowledges the financial support provided by the CSIR, New Delhi, India for CSIR Senior Research Fellowship (file no 09/796(0072)/2017-EMR-1). The research of Ng. K. Francis is funded by DST-SERB, India under Grant no. EMR/2015/001683. Bikash Thapa acknowledges the Department of Science and Technology (DST), Government of India for INSPIRE Fellowship vide Grant no. DST/INSPIRE/2018/IF180588. Animesh Barman is thankful to Bichitra Bijay Boruah for fruitful discussions.
\end{acknowledgments}

\appendix\section{}\label{sec.foo}
\begin{table}[h]
    \centering
    \begin{tabular}{c c c c c c c c c c c c  c | c c c}
    \hline
       \textrm{Field}  &  $l$ & $e^c$ & $\mu^c$ & $\tau^c$ & $h_u$ & $h_d$ & $\Phi_S$ & $\Phi_T$ & $\xi$ & $\xi^\prime$ & $\xi^{\prime\prime}$ & $\rho$ & $\phi^T_0$ & $\phi^S_0$ & $\xi_0$ \\
     \hline

     \textrm{SU(2)}  &  2 & 1 & 1 & 1 & 2 & 2 & 1 & 1 & 1 & 1 & 1 & 1 & - & - & -  \\
     
     \textrm{A}$_4$  &  3 & 1 & $1^{\prime\prime}$ & $1^\prime$ & 1 & 1 & 3 & 3 & 1 & $1^\prime$ & $1^{\prime\prime}$ & 1 & 3 & 3 &1 \\
     
    \textrm{Z}$_3$  &  $\omega$ & $\omega^2$ & $\omega^2$ & $\omega^2$ & 1 & 1 & $\omega$ & 1  & $\omega$  & $\omega$  & $\omega$  & 1 & 1 & $\omega$ & $\omega$ \\
    
    \textrm{Z}$_2$  &  l & 1 & 1 & 1 & 1 & 1 & 1 & 1 & 1 & 1 & 1 & -1 & 1 & 1 &1\\
    \hline
    \end{tabular}
    \caption{\scriptsize Full particle content of our model with associated ``\textit{driving fields}" $\phi^T_0$, $\phi^S_0$ and $\xi_0$.}
    \label{tab:2}
\end{table}

The superpotential of the model with the ``driving fields" $\phi^T_0$, $\phi^S_0$ and $\xi_0$ that allows to build the scalar potentials in the symmetry breaking sector, reads as
\begin{align}
   \nonumber W= &M (\phi^T_0 \phi_T) + g (\phi^T_0 \phi_T \phi_T) + g_1 (\phi^S_0 \phi_S \phi_S) + g_2 \xi (\phi^S_0 \phi_S)+ g_3 \xi^\prime (\xi^S_0 \xi_S)\\ &+ g_4 \xi^{\prime \prime} (\xi^S_0 \xi_S) + g_5 \xi_0 (\phi_S \phi_S) + g_6 \xi_0 \xi^2 + g_7 \xi_0 \xi^\prime \xi^{\prime\prime}
\end{align}
At this level there is no fundamental distinction among the singlets $\xi$, $\xi^\prime$ and $\xi^{\prime \prime}$. So, we can consider $\phi^S_0 \phi_S$ is coupling with $\xi$ only.\\

%$\frac{\delta W}{\delta \phi^s_{01}}= g_1. \frac{1}{3} (2 \phi^1_s \phi^1_s -\phi^2_s \phi^3_s -\phi^3_s \phi^2_s)$ + $g_2 \xi \phi^1_s + g_3 \xi^\prime \phi^3_s +g_4 \xi^\prime ^\prime\phi^2_s$ %

%$\frac{\delta W}{\delta \phi^s_{02}}= g_1. \frac{1}{3} (2 \phi^2_s \phi^2_s -\phi^1_s \phi^3_s -\phi^3_s \phi^1_s)$ + $g_2 \xi \phi^3_s + g_3 \xi^\prime \phi^2_s +g_4 \xi^\prime ^\prime\phi^1_s$ %

%$\frac{\delta W}{\delta \phi^s_{03}}= g_1. \frac{1}{3} (2 \phi^3_s \phi^3_s -\phi^1_s \phi^2_s -\phi^2_s \phi^1_s)$ + $g_2 \xi \phi^2_s + g_3 \xi^\prime \phi^1_s +g_4 \xi^\prime ^\prime\phi^3_s$

%$\frac{\delta W}{\delta \xi_0}= g_5 (\phi^1_s \phi^1_s + \phi^2_s \phi^3_s+ \phi^3_s \phi^2_s) + g_6 \xi^2 + g_7 \xi^\prime \xi^\prime ^\prime$

We use,
$$\phi_T= [\phi^1_T, \phi^2_T, \phi^3_T],\quad
\phi^T_0= [\phi^T_{01}, \phi^T_{02}, \phi^T_{03}]$$
$$\phi_S= [\phi^1_S, \phi^2_S, \phi^3_S],\quad
\phi^S_0= [\phi^S_{01}, \phi^S_{02}, \phi^S_{03}]$$

Now,
\begin{align}
\frac{\delta W}{\delta \phi^T_{01}} &= M \phi^1_T +g. \frac{1}{3}. [{2\phi^1_T \phi^1_T - \phi^2_T \phi^3_T - \phi^3_T \phi^2_T}]=0\\
\frac{\delta W}{\delta \phi^T_{02}} &= M \phi^3_T +g. \frac{1}{3}. [{2\phi^3_T \phi^3_T - \phi^1_T \phi^2_T - \phi^2_T \phi^1_T}]=0\\
\frac{\delta W}{\delta \phi^T_{03}} &= M \phi^2_T +g. \frac{1}{3}. [{2\phi^2_T \phi^2_T - \phi^1_T \phi^3_T - \phi^3_T \phi^1_T}]=0\\
\nonumber & \\
\frac{\delta W}{\delta \phi^S_{01}} &= g_1. \frac{1}{3} [2 \phi^1_S \phi^1_S -\phi^2_S \phi^3_S -\phi^3_S \phi^2_S] + g_2 \xi \phi^1_S  =0\\
\frac{\delta W}{\delta \phi^S_{02}} &= g_1. \frac{1}{3} [2 \phi^2_S \phi^2_S -\phi^1_S \phi^3_S -\phi^3_S \phi^1_S] + g_2 \xi \phi^3_S  =0\\
\frac{\delta W}{\delta \phi^S_{03}} &= g_1. \frac{1}{3} [2 \phi^3_S \phi^3_S -\phi^1_S \phi^2_S -\phi^2_S \phi^1_S] + g_2 \xi \phi^2_S =0\\
\nonumber & \\
\frac{\delta W}{\delta \xi_0} &= g_5. [\phi ^1_S \phi^1_S + \phi ^2_S \phi^3_S + \phi^3_S \phi^2_S] + g_6 \xi^2 +g_7 \xi^\prime \xi^{\prime \prime} =0    
\end{align}
A solution of the first three equation is: $$\phi_T= (v_T,0,0), v_T= -\frac{3M}{2g}$$ When we enforce $<\xi>=0$, in a finite portion of the parameter space, we find the solution as:
$$
<\xi> =0,\quad
<\xi^\prime> = u^\prime, \quad
<\xi^{\prime \prime}> = u^{\prime \prime }$$
\begin{align*}
\phi_S &= (v_S, v_S, v_S)\\
v^2_S &= -\frac{g_7 u^\prime u^{\prime \prime}}{3g_5}
\end{align*}

\bibliography{references}% Produces the bibliography via BibTeX.

\begin{thebibliography}{79}
\expandafter\ifx\csname natexlab\endcsname\relax\def\natexlab#1{#1}\fi
\expandafter\ifx\csname bibnamefont\endcsname\relax
  \def\bibnamefont#1{#1}\fi
\expandafter\ifx\csname bibfnamefont\endcsname\relax
  \def\bibfnamefont#1{#1}\fi
\expandafter\ifx\csname citenamefont\endcsname\relax
  \def\citenamefont#1{#1}\fi
\expandafter\ifx\csname url\endcsname\relax
  \def\url#1{\texttt{#1}}\fi
\expandafter\ifx\csname urlprefix\endcsname\relax\def\urlprefix{URL }\fi
\providecommand{\bibinfo}[2]{#2}
\providecommand{\eprint}[2][]{\url{#2}}

\bibitem[{\citenamefont{King et~al.}(2014)\citenamefont{King, Merle, Morisi,
  Shimizu, and Tanimoto}}]{King:2014nza}
\bibinfo{author}{\bibfnamefont{S.~F.} \bibnamefont{King}},
  \bibinfo{author}{\bibfnamefont{A.}~\bibnamefont{Merle}},
  \bibinfo{author}{\bibfnamefont{S.}~\bibnamefont{Morisi}},
  \bibinfo{author}{\bibfnamefont{Y.}~\bibnamefont{Shimizu}}, \bibnamefont{and}
  \bibinfo{author}{\bibfnamefont{M.}~\bibnamefont{Tanimoto}},
  \bibinfo{journal}{New J. Phys.} \textbf{\bibinfo{volume}{16}},
  \bibinfo{pages}{045018} (\bibinfo{year}{2014}), \eprint{1402.4271}.

\bibitem[{\citenamefont{King}(2004)}]{King:2003jb}
\bibinfo{author}{\bibfnamefont{S.~F.} \bibnamefont{King}},
  \bibinfo{journal}{Rept. Prog. Phys.} \textbf{\bibinfo{volume}{67}},
  \bibinfo{pages}{107} (\bibinfo{year}{2004}), \eprint{hep-ph/0310204}.

\bibitem[{\citenamefont{Cao et~al.}(2021)\citenamefont{Cao, Nath, Ngoc, Quyen,
  Hong~Van, and Francis}}]{Cao:2020ans}
\bibinfo{author}{\bibfnamefont{S.}~\bibnamefont{Cao}},
  \bibinfo{author}{\bibfnamefont{A.}~\bibnamefont{Nath}},
  \bibinfo{author}{\bibfnamefont{T.~V.} \bibnamefont{Ngoc}},
  \bibinfo{author}{\bibfnamefont{P.~T.} \bibnamefont{Quyen}},
  \bibinfo{author}{\bibfnamefont{N.~T.} \bibnamefont{Hong~Van}},
  \bibnamefont{and} \bibinfo{author}{\bibfnamefont{N.~K.}
  \bibnamefont{Francis}}, \bibinfo{journal}{Phys. Rev. D}
  \textbf{\bibinfo{volume}{103}}, \bibinfo{pages}{112010}
  (\bibinfo{year}{2021}), \eprint{2009.08585}.

\bibitem[{\citenamefont{McDonald}(2016)}]{McDonald:2016ixn}
\bibinfo{author}{\bibfnamefont{A.~B.} \bibnamefont{McDonald}},
  \bibinfo{journal}{Rev. Mod. Phys.} \textbf{\bibinfo{volume}{88}},
  \bibinfo{pages}{030502} (\bibinfo{year}{2016}).

\bibitem[{\citenamefont{Kajita}(2016)}]{Kajita:2016cak}
\bibinfo{author}{\bibfnamefont{T.}~\bibnamefont{Kajita}},
  \bibinfo{journal}{Rev. Mod. Phys.} \textbf{\bibinfo{volume}{88}},
  \bibinfo{pages}{030501} (\bibinfo{year}{2016}).

\bibitem[{\citenamefont{Vien}(2020)}]{Vien:2020dlk}
\bibinfo{author}{\bibfnamefont{V.~V.} \bibnamefont{Vien}},
  \bibinfo{journal}{Mod. Phys. Lett. A} \textbf{\bibinfo{volume}{35}},
  \bibinfo{pages}{2050311} (\bibinfo{year}{2020}).

\bibitem[{\citenamefont{de~Salas et~al.}(2021)\citenamefont{de~Salas, Forero,
  Gariazzo, Mart\'\i{}nez-Mirav\'e, Mena, Ternes, T\'ortola, and
  Valle}}]{deSalas:2020pgw}
\bibinfo{author}{\bibfnamefont{P.~F.} \bibnamefont{de~Salas}},
  \bibinfo{author}{\bibfnamefont{D.~V.} \bibnamefont{Forero}},
  \bibinfo{author}{\bibfnamefont{S.}~\bibnamefont{Gariazzo}},
  \bibinfo{author}{\bibfnamefont{P.}~\bibnamefont{Mart\'\i{}nez-Mirav\'e}},
  \bibinfo{author}{\bibfnamefont{O.}~\bibnamefont{Mena}},
  \bibinfo{author}{\bibfnamefont{C.~A.} \bibnamefont{Ternes}},
  \bibinfo{author}{\bibfnamefont{M.}~\bibnamefont{T\'ortola}},
  \bibnamefont{and} \bibinfo{author}{\bibfnamefont{J.~W.~F.}
  \bibnamefont{Valle}}, \bibinfo{journal}{JHEP} \textbf{\bibinfo{volume}{02}},
  \bibinfo{pages}{071} (\bibinfo{year}{2021}), \eprint{2006.11237}.

\bibitem[{\citenamefont{King}(2017)}]{King:2017guk}
\bibinfo{author}{\bibfnamefont{S.~F.} \bibnamefont{King}},
  \bibinfo{journal}{Prog. Part. Nucl. Phys.} \textbf{\bibinfo{volume}{94}},
  \bibinfo{pages}{217} (\bibinfo{year}{2017}), \eprint{1701.04413}.

\bibitem[{\citenamefont{Phong~Nguyen et~al.}(2020)\citenamefont{Phong~Nguyen,
  Hue, Si, and Thuc}}]{PhongNguyen:2017meq}
\bibinfo{author}{\bibfnamefont{T.}~\bibnamefont{Phong~Nguyen}},
  \bibinfo{author}{\bibfnamefont{L.~T.} \bibnamefont{Hue}},
  \bibinfo{author}{\bibfnamefont{D.~T.} \bibnamefont{Si}}, \bibnamefont{and}
  \bibinfo{author}{\bibfnamefont{T.~T.} \bibnamefont{Thuc}},
  \bibinfo{journal}{PTEP} \textbf{\bibinfo{volume}{2020}},
  \bibinfo{pages}{033B04} (\bibinfo{year}{2020}), \eprint{1711.05588}.

\bibitem[{\citenamefont{Nguyen et~al.}(2022)\citenamefont{Nguyen, Thuc, Si,
  Hong, and Hue}}]{Nguyen:2020ehj}
\bibinfo{author}{\bibfnamefont{T.~P.} \bibnamefont{Nguyen}},
  \bibinfo{author}{\bibfnamefont{T.~T.} \bibnamefont{Thuc}},
  \bibinfo{author}{\bibfnamefont{D.~T.} \bibnamefont{Si}},
  \bibinfo{author}{\bibfnamefont{T.~T.} \bibnamefont{Hong}}, \bibnamefont{and}
  \bibinfo{author}{\bibfnamefont{L.~T.} \bibnamefont{Hue}},
  \bibinfo{journal}{PTEP} \textbf{\bibinfo{volume}{2022}},
  \bibinfo{pages}{023B01} (\bibinfo{year}{2022}), \eprint{2011.12181}.

\bibitem[{\citenamefont{An et~al.}(2012)}]{DayaBay:2012fng}
\bibinfo{author}{\bibfnamefont{F.~P.} \bibnamefont{An}} \bibnamefont{et~al.}
  (\bibinfo{collaboration}{Daya Bay}), \bibinfo{journal}{Phys. Rev. Lett.}
  \textbf{\bibinfo{volume}{108}}, \bibinfo{pages}{171803}
  (\bibinfo{year}{2012}), \eprint{1203.1669}.

\bibitem[{\citenamefont{Ahn et~al.}(2012)}]{RENO:2012mkc}
\bibinfo{author}{\bibfnamefont{J.~K.} \bibnamefont{Ahn}} \bibnamefont{et~al.}
  (\bibinfo{collaboration}{RENO}), \bibinfo{journal}{Phys. Rev. Lett.}
  \textbf{\bibinfo{volume}{108}}, \bibinfo{pages}{191802}
  (\bibinfo{year}{2012}), \eprint{1204.0626}.

\bibitem[{\citenamefont{Adamson et~al.}(2011)}]{MINOS:2011amj}
\bibinfo{author}{\bibfnamefont{P.}~\bibnamefont{Adamson}} \bibnamefont{et~al.}
  (\bibinfo{collaboration}{MINOS}), \bibinfo{journal}{Phys. Rev. Lett.}
  \textbf{\bibinfo{volume}{107}}, \bibinfo{pages}{181802}
  (\bibinfo{year}{2011}), \eprint{1108.0015}.

\bibitem[{\citenamefont{Abe et~al.}(2012)}]{DoubleChooz:2011ymz}
\bibinfo{author}{\bibfnamefont{Y.}~\bibnamefont{Abe}} \bibnamefont{et~al.}
  (\bibinfo{collaboration}{Double Chooz}), \bibinfo{journal}{Phys. Rev. Lett.}
  \textbf{\bibinfo{volume}{108}}, \bibinfo{pages}{131801}
  (\bibinfo{year}{2012}), \eprint{1112.6353}.

\bibitem[{\citenamefont{Abe et~al.}(2011)}]{T2K:2011ypd}
\bibinfo{author}{\bibfnamefont{K.}~\bibnamefont{Abe}} \bibnamefont{et~al.}
  (\bibinfo{collaboration}{T2K}), \bibinfo{journal}{Phys. Rev. Lett.}
  \textbf{\bibinfo{volume}{107}}, \bibinfo{pages}{041801}
  (\bibinfo{year}{2011}), \eprint{1106.2822}.

\bibitem[{\citenamefont{Furry}(1939)}]{Furry:1939qr}
\bibinfo{author}{\bibfnamefont{W.~H.} \bibnamefont{Furry}},
  \bibinfo{journal}{Phys. Rev.} \textbf{\bibinfo{volume}{56}},
  \bibinfo{pages}{1184} (\bibinfo{year}{1939}).

\bibitem[{\citenamefont{Dell'Oro et~al.}(2016)\citenamefont{Dell'Oro, Marcocci,
  Viel, and Vissani}}]{DellOro:2016tmg}
\bibinfo{author}{\bibfnamefont{S.}~\bibnamefont{Dell'Oro}},
  \bibinfo{author}{\bibfnamefont{S.}~\bibnamefont{Marcocci}},
  \bibinfo{author}{\bibfnamefont{M.}~\bibnamefont{Viel}}, \bibnamefont{and}
  \bibinfo{author}{\bibfnamefont{F.}~\bibnamefont{Vissani}},
  \bibinfo{journal}{Adv. High Energy Phys.} \textbf{\bibinfo{volume}{2016}},
  \bibinfo{pages}{2162659} (\bibinfo{year}{2016}), \eprint{1601.07512}.

\bibitem[{\citenamefont{Bilenky and Giunti}(2012)}]{Bilenky:2012qi}
\bibinfo{author}{\bibfnamefont{S.~M.} \bibnamefont{Bilenky}} \bibnamefont{and}
  \bibinfo{author}{\bibfnamefont{C.}~\bibnamefont{Giunti}},
  \bibinfo{journal}{Mod. Phys. Lett. A} \textbf{\bibinfo{volume}{27}},
  \bibinfo{pages}{1230015} (\bibinfo{year}{2012}), \eprint{1203.5250}.

\bibitem[{\citenamefont{Weinberg}(1979)}]{Weinberg:1979sa}
\bibinfo{author}{\bibfnamefont{S.}~\bibnamefont{Weinberg}},
  \bibinfo{journal}{Phys. Rev. Lett.} \textbf{\bibinfo{volume}{43}},
  \bibinfo{pages}{1566} (\bibinfo{year}{1979}).

\bibitem[{\citenamefont{Yanagida}(1980)}]{Yanagida:1980xy}
\bibinfo{author}{\bibfnamefont{T.}~\bibnamefont{Yanagida}},
  \bibinfo{journal}{Prog. Theor. Phys.} \textbf{\bibinfo{volume}{64}},
  \bibinfo{pages}{1103} (\bibinfo{year}{1980}).

\bibitem[{\citenamefont{Minkowski}(1977)}]{Minkowski:1977sc}
\bibinfo{author}{\bibfnamefont{P.}~\bibnamefont{Minkowski}},
  \bibinfo{journal}{Phys. Lett. B} \textbf{\bibinfo{volume}{67}},
  \bibinfo{pages}{421} (\bibinfo{year}{1977}).

\bibitem[{\citenamefont{Das and Das}(2020)}]{Das:2019kmn}
\bibinfo{author}{\bibfnamefont{P.}~\bibnamefont{Das}} \bibnamefont{and}
  \bibinfo{author}{\bibfnamefont{M.~K.} \bibnamefont{Das}},
  \bibinfo{journal}{Int. J. Mod. Phys. A} \textbf{\bibinfo{volume}{35}},
  \bibinfo{pages}{2050125} (\bibinfo{year}{2020}), \eprint{1908.08417}.

\bibitem[{\citenamefont{Gell-Mann}(1979)}]{gell1979ramond}
\bibinfo{author}{\bibfnamefont{M.}~\bibnamefont{Gell-Mann}},
  \bibinfo{journal}{Supergravity, P. van Nieuwenhuizen and DZ Freedman eds.,
  North Holland, Amsterdam The Netherlands} \textbf{\bibinfo{volume}{315}}
  (\bibinfo{year}{1979}).

\bibitem[{\citenamefont{Mohapatra and Senjanovic}(1980)}]{Mohapatra:1979ia}
\bibinfo{author}{\bibfnamefont{R.~N.} \bibnamefont{Mohapatra}}
  \bibnamefont{and}
  \bibinfo{author}{\bibfnamefont{G.}~\bibnamefont{Senjanovic}},
  \bibinfo{journal}{Phys. Rev. Lett.} \textbf{\bibinfo{volume}{44}},
  \bibinfo{pages}{912} (\bibinfo{year}{1980}).

\bibitem[{\citenamefont{Fukuyama and Nishiura}(1997)}]{Fukuyama:1997ky}
\bibinfo{author}{\bibfnamefont{T.}~\bibnamefont{Fukuyama}} \bibnamefont{and}
  \bibinfo{author}{\bibfnamefont{H.}~\bibnamefont{Nishiura}}
  (\bibinfo{year}{1997}), \eprint{hep-ph/9702253}.

\bibitem[{\citenamefont{Vien et~al.}(2019)\citenamefont{Vien, Long, and
  C\'arcamo~Hern\'andez}}]{Vien:2019zhs}
\bibinfo{author}{\bibfnamefont{V.~V.} \bibnamefont{Vien}},
  \bibinfo{author}{\bibfnamefont{H.~N.} \bibnamefont{Long}}, \bibnamefont{and}
  \bibinfo{author}{\bibfnamefont{A.~E.} \bibnamefont{C\'arcamo~Hern\'andez}},
  \bibinfo{journal}{PTEP} \textbf{\bibinfo{volume}{2019}},
  \bibinfo{pages}{113B04} (\bibinfo{year}{2019}), \eprint{1909.09532}.

\bibitem[{\citenamefont{Yanagida}(1979)}]{yanagida1979horizontal}
\bibinfo{author}{\bibfnamefont{T.}~\bibnamefont{Yanagida}}, in
  \emph{\bibinfo{booktitle}{Workshop on unified theory and baryon number in the
  universe}} (\bibinfo{organization}{KEK, Tsukuba, Japan},
  \bibinfo{year}{1979}).

\bibitem[{\citenamefont{Boruah and Das}(2022)}]{Boruah:2021ktk}
\bibinfo{author}{\bibfnamefont{B.~B.} \bibnamefont{Boruah}} \bibnamefont{and}
  \bibinfo{author}{\bibfnamefont{M.~K.} \bibnamefont{Das}},
  \bibinfo{journal}{Int. J. Mod. Phys. A} \textbf{\bibinfo{volume}{37}},
  \bibinfo{pages}{2250026} (\bibinfo{year}{2022}), \eprint{2111.10341}.

\bibitem[{\citenamefont{Ma}(1998)}]{Ma:1998ias}
\bibinfo{author}{\bibfnamefont{E.}~\bibnamefont{Ma}}, \bibinfo{journal}{PoS}
  \textbf{\bibinfo{volume}{corfu98}}, \bibinfo{pages}{047}
  (\bibinfo{year}{1998}), \eprint{hep-ph/9902450}.

\bibitem[{\citenamefont{Csaki}(1996)}]{Csaki:1996ks}
\bibinfo{author}{\bibfnamefont{C.}~\bibnamefont{Csaki}}, \bibinfo{journal}{Mod.
  Phys. Lett. A} \textbf{\bibinfo{volume}{11}}, \bibinfo{pages}{599}
  (\bibinfo{year}{1996}), \eprint{hep-ph/9606414}.

\bibitem[{\citenamefont{Ellwanger et~al.}(2010)\citenamefont{Ellwanger,
  Hugonie, and Teixeira}}]{Ellwanger:2009dp}
\bibinfo{author}{\bibfnamefont{U.}~\bibnamefont{Ellwanger}},
  \bibinfo{author}{\bibfnamefont{C.}~\bibnamefont{Hugonie}}, \bibnamefont{and}
  \bibinfo{author}{\bibfnamefont{A.~M.} \bibnamefont{Teixeira}},
  \bibinfo{journal}{Phys. Rept.} \textbf{\bibinfo{volume}{496}},
  \bibinfo{pages}{1} (\bibinfo{year}{2010}), \eprint{0910.1785}.

\bibitem[{\citenamefont{Ibanez and Uranga}(2012)}]{Ibanez:2012zz}
\bibinfo{author}{\bibfnamefont{L.~E.} \bibnamefont{Ibanez}} \bibnamefont{and}
  \bibinfo{author}{\bibfnamefont{A.~M.} \bibnamefont{Uranga}},
  \emph{\bibinfo{title}{{String theory and particle physics: An introduction to
  string phenomenology}}} (\bibinfo{publisher}{Cambridge University Press},
  \bibinfo{year}{2012}), ISBN \bibinfo{isbn}{978-0-521-51752-2,
  978-1-139-22742-1}.

\bibitem[{\citenamefont{Arkani-Hamed et~al.}(2001)\citenamefont{Arkani-Hamed,
  Dimopoulos, Dvali, and March-Russell}}]{Arkani-Hamed:1998wuz}
\bibinfo{author}{\bibfnamefont{N.}~\bibnamefont{Arkani-Hamed}},
  \bibinfo{author}{\bibfnamefont{S.}~\bibnamefont{Dimopoulos}},
  \bibinfo{author}{\bibfnamefont{G.~R.} \bibnamefont{Dvali}}, \bibnamefont{and}
  \bibinfo{author}{\bibfnamefont{J.}~\bibnamefont{March-Russell}},
  \bibinfo{journal}{Phys. Rev. D} \textbf{\bibinfo{volume}{65}},
  \bibinfo{pages}{024032} (\bibinfo{year}{2001}), \eprint{hep-ph/9811448}.

\bibitem[{\citenamefont{Ma}(2006{\natexlab{a}})}]{Ma:2006km}
\bibinfo{author}{\bibfnamefont{E.}~\bibnamefont{Ma}}, \bibinfo{journal}{Phys.
  Rev. D} \textbf{\bibinfo{volume}{73}}, \bibinfo{pages}{077301}
  (\bibinfo{year}{2006}{\natexlab{a}}), \eprint{hep-ph/0601225}.

\bibitem[{\citenamefont{Aker et~al.}(2019)\citenamefont{Aker, Altenm{\"u}ller,
  Arenz, Babutzka, Barrett, Bauer, Beck, Beglarian, Behrens, Bergmann
  et~al.}}]{aker2019improved}
\bibinfo{author}{\bibfnamefont{M.}~\bibnamefont{Aker}},
  \bibinfo{author}{\bibfnamefont{K.}~\bibnamefont{Altenm{\"u}ller}},
  \bibinfo{author}{\bibfnamefont{M.}~\bibnamefont{Arenz}},
  \bibinfo{author}{\bibfnamefont{M.}~\bibnamefont{Babutzka}},
  \bibinfo{author}{\bibfnamefont{J.}~\bibnamefont{Barrett}},
  \bibinfo{author}{\bibfnamefont{S.}~\bibnamefont{Bauer}},
  \bibinfo{author}{\bibfnamefont{M.}~\bibnamefont{Beck}},
  \bibinfo{author}{\bibfnamefont{A.}~\bibnamefont{Beglarian}},
  \bibinfo{author}{\bibfnamefont{J.}~\bibnamefont{Behrens}},
  \bibinfo{author}{\bibfnamefont{T.}~\bibnamefont{Bergmann}},
  \bibnamefont{et~al.}, \bibinfo{journal}{Physical review letters}
  \textbf{\bibinfo{volume}{123}}, \bibinfo{pages}{221802}
  (\bibinfo{year}{2019}).

\bibitem[{\citenamefont{Francis}(2014)}]{Francis:2014dya}
\bibinfo{author}{\bibfnamefont{N.~K.} \bibnamefont{Francis}},
  \bibinfo{journal}{Adv. High Energy Phys.} \textbf{\bibinfo{volume}{2014}},
  \bibinfo{pages}{689719} (\bibinfo{year}{2014}).

\bibitem[{\citenamefont{Group et~al.}(2020)\citenamefont{Group, Zyla, Barnett,
  Beringer, Dahl, Dwyer, Groom, Lin, Lugovsky, Pianori
  et~al.}}]{particle2020review}
\bibinfo{author}{\bibfnamefont{P.~D.} \bibnamefont{Group}},
  \bibinfo{author}{\bibfnamefont{P.}~\bibnamefont{Zyla}},
  \bibinfo{author}{\bibfnamefont{R.}~\bibnamefont{Barnett}},
  \bibinfo{author}{\bibfnamefont{J.}~\bibnamefont{Beringer}},
  \bibinfo{author}{\bibfnamefont{O.}~\bibnamefont{Dahl}},
  \bibinfo{author}{\bibfnamefont{D.}~\bibnamefont{Dwyer}},
  \bibinfo{author}{\bibfnamefont{D.}~\bibnamefont{Groom}},
  \bibinfo{author}{\bibfnamefont{C.-J.} \bibnamefont{Lin}},
  \bibinfo{author}{\bibfnamefont{K.}~\bibnamefont{Lugovsky}},
  \bibinfo{author}{\bibfnamefont{E.}~\bibnamefont{Pianori}},
  \bibnamefont{et~al.}, \bibinfo{journal}{Progress of Theoretical and
  Experimental Physics} \textbf{\bibinfo{volume}{2020}},
  \bibinfo{pages}{083C01} (\bibinfo{year}{2020}).

\bibitem[{\citenamefont{Nath and Francis}(2021)}]{Nath:2018ywc}
\bibinfo{author}{\bibfnamefont{A.}~\bibnamefont{Nath}} \bibnamefont{and}
  \bibinfo{author}{\bibfnamefont{N.~K.} \bibnamefont{Francis}},
  \bibinfo{journal}{Int. J. Mod. Phys. A} \textbf{\bibinfo{volume}{36}},
  \bibinfo{pages}{2130008} (\bibinfo{year}{2021}), \eprint{1804.08467}.

\bibitem[{\citenamefont{Harrison et~al.}(2002)\citenamefont{Harrison, Perkins,
  and Scott}}]{Harrison:2002er}
\bibinfo{author}{\bibfnamefont{P.~F.} \bibnamefont{Harrison}},
  \bibinfo{author}{\bibfnamefont{D.~H.} \bibnamefont{Perkins}},
  \bibnamefont{and} \bibinfo{author}{\bibfnamefont{W.~G.} \bibnamefont{Scott}},
  \bibinfo{journal}{Phys. Lett. B} \textbf{\bibinfo{volume}{530}},
  \bibinfo{pages}{167} (\bibinfo{year}{2002}), \eprint{hep-ph/0202074}.

\bibitem[{\citenamefont{Harrison and Scott}(2002)}]{Harrison:2002kp}
\bibinfo{author}{\bibfnamefont{P.~F.} \bibnamefont{Harrison}} \bibnamefont{and}
  \bibinfo{author}{\bibfnamefont{W.~G.} \bibnamefont{Scott}},
  \bibinfo{journal}{Phys. Lett. B} \textbf{\bibinfo{volume}{535}},
  \bibinfo{pages}{163} (\bibinfo{year}{2002}), \eprint{hep-ph/0203209}.

\bibitem[{\citenamefont{Albright and Rodejohann}(2009)}]{Albright:2008rp}
\bibinfo{author}{\bibfnamefont{C.~H.} \bibnamefont{Albright}} \bibnamefont{and}
  \bibinfo{author}{\bibfnamefont{W.}~\bibnamefont{Rodejohann}},
  \bibinfo{journal}{Eur. Phys. J. C} \textbf{\bibinfo{volume}{62}},
  \bibinfo{pages}{599} (\bibinfo{year}{2009}), \eprint{0812.0436}.

\bibitem[{\citenamefont{He and Zee}(2011)}]{He:2011gb}
\bibinfo{author}{\bibfnamefont{X.-G.} \bibnamefont{He}} \bibnamefont{and}
  \bibinfo{author}{\bibfnamefont{A.}~\bibnamefont{Zee}},
  \bibinfo{journal}{Phys. Rev. D} \textbf{\bibinfo{volume}{84}},
  \bibinfo{pages}{053004} (\bibinfo{year}{2011}), \eprint{1106.4359}.

\bibitem[{\citenamefont{Thapa and Francis}(2021)}]{thapa2021resonant}
\bibinfo{author}{\bibfnamefont{B.}~\bibnamefont{Thapa}} \bibnamefont{and}
  \bibinfo{author}{\bibfnamefont{N.~K.} \bibnamefont{Francis}},
  \bibinfo{journal}{The European Physical Journal C}
  \textbf{\bibinfo{volume}{81}}, \bibinfo{pages}{1} (\bibinfo{year}{2021}).

\bibitem[{\citenamefont{Francis and Singh}(2012)}]{Francis:2012jj}
\bibinfo{author}{\bibfnamefont{N.~K.} \bibnamefont{Francis}} \bibnamefont{and}
  \bibinfo{author}{\bibfnamefont{N.~N.} \bibnamefont{Singh}},
  \bibinfo{journal}{Nucl. Phys. B} \textbf{\bibinfo{volume}{863}},
  \bibinfo{pages}{19} (\bibinfo{year}{2012}), \eprint{1206.3420}.

\bibitem[{\citenamefont{Boucenna et~al.}(2012)\citenamefont{Boucenna, Morisi,
  Tortola, and Valle}}]{Boucenna:2012xb}
\bibinfo{author}{\bibfnamefont{S.~M.} \bibnamefont{Boucenna}},
  \bibinfo{author}{\bibfnamefont{S.}~\bibnamefont{Morisi}},
  \bibinfo{author}{\bibfnamefont{M.}~\bibnamefont{Tortola}}, \bibnamefont{and}
  \bibinfo{author}{\bibfnamefont{J.~W.~F.} \bibnamefont{Valle}},
  \bibinfo{journal}{Phys. Rev. D} \textbf{\bibinfo{volume}{86}},
  \bibinfo{pages}{051301} (\bibinfo{year}{2012}), \eprint{1206.2555}.

\bibitem[{\citenamefont{Chen et~al.}(2019)\citenamefont{Chen, Ding, Srivastava,
  and Valle}}]{Chen:2019egu}
\bibinfo{author}{\bibfnamefont{P.}~\bibnamefont{Chen}},
  \bibinfo{author}{\bibfnamefont{G.-J.} \bibnamefont{Ding}},
  \bibinfo{author}{\bibfnamefont{R.}~\bibnamefont{Srivastava}},
  \bibnamefont{and} \bibinfo{author}{\bibfnamefont{J.~W.~F.}
  \bibnamefont{Valle}}, \bibinfo{journal}{Phys. Lett. B}
  \textbf{\bibinfo{volume}{792}}, \bibinfo{pages}{461} (\bibinfo{year}{2019}),
  \eprint{1902.08962}.

\bibitem[{\citenamefont{Ding et~al.}(2019)\citenamefont{Ding, Nath, Srivastava,
  and Valle}}]{Ding:2019vvi}
\bibinfo{author}{\bibfnamefont{G.-J.} \bibnamefont{Ding}},
  \bibinfo{author}{\bibfnamefont{N.}~\bibnamefont{Nath}},
  \bibinfo{author}{\bibfnamefont{R.}~\bibnamefont{Srivastava}},
  \bibnamefont{and} \bibinfo{author}{\bibfnamefont{J.~W.~F.}
  \bibnamefont{Valle}}, \bibinfo{journal}{Phys. Lett. B}
  \textbf{\bibinfo{volume}{796}}, \bibinfo{pages}{162} (\bibinfo{year}{2019}),
  \eprint{1904.05632}.

\bibitem[{\citenamefont{King and Luhn}(2013)}]{King:2013eh}
\bibinfo{author}{\bibfnamefont{S.~F.} \bibnamefont{King}} \bibnamefont{and}
  \bibinfo{author}{\bibfnamefont{C.}~\bibnamefont{Luhn}},
  \bibinfo{journal}{Rept. Prog. Phys.} \textbf{\bibinfo{volume}{76}},
  \bibinfo{pages}{056201} (\bibinfo{year}{2013}), \eprint{1301.1340}.

\bibitem[{\citenamefont{Ishimori et~al.}(2010)\citenamefont{Ishimori,
  Kobayashi, Ohki, Shimizu, Okada, and Tanimoto}}]{Ishimori:2010au}
\bibinfo{author}{\bibfnamefont{H.}~\bibnamefont{Ishimori}},
  \bibinfo{author}{\bibfnamefont{T.}~\bibnamefont{Kobayashi}},
  \bibinfo{author}{\bibfnamefont{H.}~\bibnamefont{Ohki}},
  \bibinfo{author}{\bibfnamefont{Y.}~\bibnamefont{Shimizu}},
  \bibinfo{author}{\bibfnamefont{H.}~\bibnamefont{Okada}}, \bibnamefont{and}
  \bibinfo{author}{\bibfnamefont{M.}~\bibnamefont{Tanimoto}},
  \bibinfo{journal}{Prog. Theor. Phys. Suppl.} \textbf{\bibinfo{volume}{183}},
  \bibinfo{pages}{1} (\bibinfo{year}{2010}), \eprint{1003.3552}.

\bibitem[{\citenamefont{Ma}(2006{\natexlab{b}})}]{Ma:2005qf}
\bibinfo{author}{\bibfnamefont{E.}~\bibnamefont{Ma}}, \bibinfo{journal}{Phys.
  Rev. D} \textbf{\bibinfo{volume}{73}}, \bibinfo{pages}{057304}
  (\bibinfo{year}{2006}{\natexlab{b}}), \eprint{hep-ph/0511133}.

\bibitem[{\citenamefont{Ma}(2016)}]{Ma:2015fpa}
\bibinfo{author}{\bibfnamefont{E.}~\bibnamefont{Ma}}, \bibinfo{journal}{Phys.
  Lett. B} \textbf{\bibinfo{volume}{752}}, \bibinfo{pages}{198}
  (\bibinfo{year}{2016}), \eprint{1510.02501}.

\bibitem[{\citenamefont{Ma}(2004)}]{Ma:2004zd}
\bibinfo{author}{\bibfnamefont{E.}~\bibnamefont{Ma}}, \bibinfo{journal}{New J.
  Phys.} \textbf{\bibinfo{volume}{6}}, \bibinfo{pages}{104}
  (\bibinfo{year}{2004}), \eprint{hep-ph/0405152}.

\bibitem[{\citenamefont{Bazzocchi and Merlo}(2013)}]{Bazzocchi:2012st}
\bibinfo{author}{\bibfnamefont{F.}~\bibnamefont{Bazzocchi}} \bibnamefont{and}
  \bibinfo{author}{\bibfnamefont{L.}~\bibnamefont{Merlo}},
  \bibinfo{journal}{Fortsch. Phys.} \textbf{\bibinfo{volume}{61}},
  \bibinfo{pages}{571} (\bibinfo{year}{2013}), \eprint{1205.5135}.

\bibitem[{\citenamefont{Ma}(2005)}]{Ma:2005mw}
\bibinfo{author}{\bibfnamefont{E.}~\bibnamefont{Ma}}, \bibinfo{journal}{Mod.
  Phys. Lett. A} \textbf{\bibinfo{volume}{20}}, \bibinfo{pages}{2601}
  (\bibinfo{year}{2005}), \eprint{hep-ph/0508099}.

\bibitem[{\citenamefont{Vien et~al.}(2015)\citenamefont{Vien, Long, and
  Khoi}}]{Vien:2015fhk}
\bibinfo{author}{\bibfnamefont{V.~V.} \bibnamefont{Vien}},
  \bibinfo{author}{\bibfnamefont{H.~N.} \bibnamefont{Long}}, \bibnamefont{and}
  \bibinfo{author}{\bibfnamefont{D.~P.} \bibnamefont{Khoi}},
  \bibinfo{journal}{Int. J. Mod. Phys. A} \textbf{\bibinfo{volume}{30}},
  \bibinfo{pages}{1550102} (\bibinfo{year}{2015}), \eprint{1506.06063}.

\bibitem[{\citenamefont{Ma}(2008)}]{ma2008near}
\bibinfo{author}{\bibfnamefont{E.}~\bibnamefont{Ma}}, \bibinfo{journal}{Physics
  Letters B} \textbf{\bibinfo{volume}{660}}, \bibinfo{pages}{505}
  (\bibinfo{year}{2008}).

\bibitem[{\citenamefont{de~Medeiros~Varzielas
  et~al.}(2007)\citenamefont{de~Medeiros~Varzielas, King, and
  Ross}}]{de2007neutrino}
\bibinfo{author}{\bibfnamefont{I.}~\bibnamefont{de~Medeiros~Varzielas}},
  \bibinfo{author}{\bibfnamefont{S.}~\bibnamefont{King}}, \bibnamefont{and}
  \bibinfo{author}{\bibfnamefont{G.}~\bibnamefont{Ross}},
  \bibinfo{journal}{Physics Letters B} \textbf{\bibinfo{volume}{648}},
  \bibinfo{pages}{201} (\bibinfo{year}{2007}).

\bibitem[{\citenamefont{Harrison et~al.}(2014)\citenamefont{Harrison, Krishnan,
  and Scott}}]{harrison2014deviations}
\bibinfo{author}{\bibfnamefont{P.}~\bibnamefont{Harrison}},
  \bibinfo{author}{\bibfnamefont{R.}~\bibnamefont{Krishnan}}, \bibnamefont{and}
  \bibinfo{author}{\bibfnamefont{W.}~\bibnamefont{Scott}},
  \bibinfo{journal}{International Journal of Modern Physics A}
  \textbf{\bibinfo{volume}{29}}, \bibinfo{pages}{1450095}
  (\bibinfo{year}{2014}).

\bibitem[{\citenamefont{Ishimori et~al.}(2009)\citenamefont{Ishimori,
  Kobayashi, Okada, Shimizu, and Tanimoto}}]{Ishimori:2008uc}
\bibinfo{author}{\bibfnamefont{H.}~\bibnamefont{Ishimori}},
  \bibinfo{author}{\bibfnamefont{T.}~\bibnamefont{Kobayashi}},
  \bibinfo{author}{\bibfnamefont{H.}~\bibnamefont{Okada}},
  \bibinfo{author}{\bibfnamefont{Y.}~\bibnamefont{Shimizu}}, \bibnamefont{and}
  \bibinfo{author}{\bibfnamefont{M.}~\bibnamefont{Tanimoto}},
  \bibinfo{journal}{JHEP} \textbf{\bibinfo{volume}{04}}, \bibinfo{pages}{011}
  (\bibinfo{year}{2009}), \eprint{0811.4683}.

\bibitem[{\citenamefont{Loualidi}(2021)}]{loualidi2021trimaximal}
\bibinfo{author}{\bibfnamefont{M.}~\bibnamefont{Loualidi}},
  \bibinfo{journal}{arXiv preprint arXiv:2104.13734}  (\bibinfo{year}{2021}).

\bibitem[{\citenamefont{Altarelli and Feruglio}(2006)}]{Altarelli:2005yx}
\bibinfo{author}{\bibfnamefont{G.}~\bibnamefont{Altarelli}} \bibnamefont{and}
  \bibinfo{author}{\bibfnamefont{F.}~\bibnamefont{Feruglio}},
  \bibinfo{journal}{Nucl. Phys. B} \textbf{\bibinfo{volume}{741}},
  \bibinfo{pages}{215} (\bibinfo{year}{2006}), \eprint{hep-ph/0512103}.

\bibitem[{\citenamefont{Altarelli and Feruglio}(2005)}]{Altarelli:2005yp}
\bibinfo{author}{\bibfnamefont{G.}~\bibnamefont{Altarelli}} \bibnamefont{and}
  \bibinfo{author}{\bibfnamefont{F.}~\bibnamefont{Feruglio}},
  \bibinfo{journal}{Nucl. Phys. B} \textbf{\bibinfo{volume}{720}},
  \bibinfo{pages}{64} (\bibinfo{year}{2005}), \eprint{hep-ph/0504165}.

\bibitem[{\citenamefont{Altarelli and Feruglio}(2010)}]{Altarelli:2010gt}
\bibinfo{author}{\bibfnamefont{G.}~\bibnamefont{Altarelli}} \bibnamefont{and}
  \bibinfo{author}{\bibfnamefont{F.}~\bibnamefont{Feruglio}},
  \bibinfo{journal}{Rev. Mod. Phys.} \textbf{\bibinfo{volume}{82}},
  \bibinfo{pages}{2701} (\bibinfo{year}{2010}), \eprint{1002.0211}.

\bibitem[{\citenamefont{Brahmachari et~al.}(2008)\citenamefont{Brahmachari,
  Choubey, and Mitra}}]{Brahmachari:2008fn}
\bibinfo{author}{\bibfnamefont{B.}~\bibnamefont{Brahmachari}},
  \bibinfo{author}{\bibfnamefont{S.}~\bibnamefont{Choubey}}, \bibnamefont{and}
  \bibinfo{author}{\bibfnamefont{M.}~\bibnamefont{Mitra}},
  \bibinfo{journal}{Phys. Rev. D} \textbf{\bibinfo{volume}{77}},
  \bibinfo{pages}{073008} (\bibinfo{year}{2008}), \bibinfo{note}{[Erratum:
  Phys.Rev.D 77, 119901 (2008)]}, \eprint{0801.3554}.

\bibitem[{\citenamefont{Shimizu et~al.}(2011)\citenamefont{Shimizu, Tanimoto,
  and Watanabe}}]{Shimizu:2011xg}
\bibinfo{author}{\bibfnamefont{Y.}~\bibnamefont{Shimizu}},
  \bibinfo{author}{\bibfnamefont{M.}~\bibnamefont{Tanimoto}}, \bibnamefont{and}
  \bibinfo{author}{\bibfnamefont{A.}~\bibnamefont{Watanabe}},
  \bibinfo{journal}{Prog. Theor. Phys.} \textbf{\bibinfo{volume}{126}},
  \bibinfo{pages}{81} (\bibinfo{year}{2011}), \eprint{1105.2929}.

\bibitem[{\citenamefont{Karmakar and Sil}(2015)}]{Karmakar:2014dva}
\bibinfo{author}{\bibfnamefont{B.}~\bibnamefont{Karmakar}} \bibnamefont{and}
  \bibinfo{author}{\bibfnamefont{A.}~\bibnamefont{Sil}},
  \bibinfo{journal}{Phys. Rev. D} \textbf{\bibinfo{volume}{91}},
  \bibinfo{pages}{013004} (\bibinfo{year}{2015}), \eprint{1407.5826}.

\bibitem[{\citenamefont{King and Luhn}(2011)}]{King:2011zj}
\bibinfo{author}{\bibfnamefont{S.~F.} \bibnamefont{King}} \bibnamefont{and}
  \bibinfo{author}{\bibfnamefont{C.}~\bibnamefont{Luhn}},
  \bibinfo{journal}{JHEP} \textbf{\bibinfo{volume}{09}}, \bibinfo{pages}{042}
  (\bibinfo{year}{2011}), \eprint{1107.5332}.

\bibitem[{\citenamefont{Cooper et~al.}(2012)\citenamefont{Cooper, King, and
  Luhn}}]{Cooper:2011rh}
\bibinfo{author}{\bibfnamefont{I.~K.} \bibnamefont{Cooper}},
  \bibinfo{author}{\bibfnamefont{S.~F.} \bibnamefont{King}}, \bibnamefont{and}
  \bibinfo{author}{\bibfnamefont{C.}~\bibnamefont{Luhn}},
  \bibinfo{journal}{Nucl. Phys. B} \textbf{\bibinfo{volume}{859}},
  \bibinfo{pages}{159} (\bibinfo{year}{2012}), \eprint{1110.5676}.

\bibitem[{\citenamefont{Ding and Meloni}(2012)}]{Ding:2011gt}
\bibinfo{author}{\bibfnamefont{G.-J.} \bibnamefont{Ding}} \bibnamefont{and}
  \bibinfo{author}{\bibfnamefont{D.}~\bibnamefont{Meloni}},
  \bibinfo{journal}{Nucl. Phys. B} \textbf{\bibinfo{volume}{855}},
  \bibinfo{pages}{21} (\bibinfo{year}{2012}), \eprint{1108.2733}.

\bibitem[{\citenamefont{Ahn and Kang}(2012)}]{Ahn:2012tv}
\bibinfo{author}{\bibfnamefont{Y.~H.} \bibnamefont{Ahn}} \bibnamefont{and}
  \bibinfo{author}{\bibfnamefont{S.~K.} \bibnamefont{Kang}},
  \bibinfo{journal}{Phys. Rev. D} \textbf{\bibinfo{volume}{86}},
  \bibinfo{pages}{093003} (\bibinfo{year}{2012}), \eprint{1203.4185}.

\bibitem[{\citenamefont{Ahn et~al.}(2013)\citenamefont{Ahn, Kang, and
  Kim}}]{Ahn:2013mva}
\bibinfo{author}{\bibfnamefont{Y.~H.} \bibnamefont{Ahn}},
  \bibinfo{author}{\bibfnamefont{S.~K.} \bibnamefont{Kang}}, \bibnamefont{and}
  \bibinfo{author}{\bibfnamefont{C.~S.} \bibnamefont{Kim}},
  \bibinfo{journal}{Phys. Rev. D} \textbf{\bibinfo{volume}{87}},
  \bibinfo{pages}{113012} (\bibinfo{year}{2013}), \eprint{1304.0921}.

\bibitem[{\citenamefont{Kang et~al.}(2018)\citenamefont{Kang, Shimizu, Takagi,
  Takahashi, and Tanimoto}}]{Kang:2018txu}
\bibinfo{author}{\bibfnamefont{S.~K.} \bibnamefont{Kang}},
  \bibinfo{author}{\bibfnamefont{Y.}~\bibnamefont{Shimizu}},
  \bibinfo{author}{\bibfnamefont{K.}~\bibnamefont{Takagi}},
  \bibinfo{author}{\bibfnamefont{S.}~\bibnamefont{Takahashi}},
  \bibnamefont{and} \bibinfo{author}{\bibfnamefont{M.}~\bibnamefont{Tanimoto}},
  \bibinfo{journal}{PTEP} \textbf{\bibinfo{volume}{2018}},
  \bibinfo{pages}{083B01} (\bibinfo{year}{2018}), \eprint{1804.10468}.

\bibitem[{\citenamefont{Esteban et~al.}(2020)\citenamefont{Esteban,
  Gonzalez-Garcia, Maltoni, Schwetz, and Zhou}}]{Esteban:2020cvm}
\bibinfo{author}{\bibfnamefont{I.}~\bibnamefont{Esteban}},
  \bibinfo{author}{\bibfnamefont{M.~C.} \bibnamefont{Gonzalez-Garcia}},
  \bibinfo{author}{\bibfnamefont{M.}~\bibnamefont{Maltoni}},
  \bibinfo{author}{\bibfnamefont{T.}~\bibnamefont{Schwetz}}, \bibnamefont{and}
  \bibinfo{author}{\bibfnamefont{A.}~\bibnamefont{Zhou}},
  \bibinfo{journal}{JHEP} \textbf{\bibinfo{volume}{09}}, \bibinfo{pages}{178}
  (\bibinfo{year}{2020}), \eprint{2007.14792}.

\bibitem[{\citenamefont{Gando}(2020)}]{Gando:2020cxo}
\bibinfo{author}{\bibfnamefont{Y.}~\bibnamefont{Gando}}
  (\bibinfo{collaboration}{KamLAND-Zen}), \bibinfo{journal}{J. Phys. Conf.
  Ser.} \textbf{\bibinfo{volume}{1468}}, \bibinfo{pages}{012142}
  (\bibinfo{year}{2020}).

\bibitem[{\citenamefont{Biller}(2021)}]{Biller:2021bqx}
\bibinfo{author}{\bibfnamefont{S.~D.} \bibnamefont{Biller}},
  \bibinfo{journal}{Phys. Rev. D} \textbf{\bibinfo{volume}{104}},
  \bibinfo{pages}{012002} (\bibinfo{year}{2021}), \eprint{2103.06036}.

\bibitem[{\citenamefont{Di~Marco}(2020)}]{DiMarco:2020mdk}
\bibinfo{author}{\bibfnamefont{N.}~\bibnamefont{Di~Marco}}
  (\bibinfo{collaboration}{GERDA}), \bibinfo{journal}{Nucl. Instrum. Meth. A}
  \textbf{\bibinfo{volume}{958}}, \bibinfo{pages}{162112}
  (\bibinfo{year}{2020}).

\bibitem[{\citenamefont{Agostini et~al.}(2020)}]{GERDA:2017wlm}
\bibinfo{author}{\bibfnamefont{M.}~\bibnamefont{Agostini}} \bibnamefont{et~al.}
  (\bibinfo{collaboration}{GERDA}), \bibinfo{journal}{J. Phys. Conf. Ser.}
  \textbf{\bibinfo{volume}{1342}}, \bibinfo{pages}{012005}
  (\bibinfo{year}{2020}), \eprint{1710.07776}.

\bibitem[{\citenamefont{D'Andrea}(2020)}]{DAndrea:2020ezp}
\bibinfo{author}{\bibfnamefont{V.}~\bibnamefont{D'Andrea}}
  (\bibinfo{collaboration}{GERDA}), \bibinfo{journal}{Nuovo Cim. C}
  \textbf{\bibinfo{volume}{43}}, \bibinfo{pages}{24} (\bibinfo{year}{2020}).

\bibitem[{\citenamefont{Abgrall et~al.}(2021)\citenamefont{Abgrall, Abt,
  Agostini, Alexander, Andreoiu, Araujo, Avignone~III, Bae, Bakalyarov, Balata
  et~al.}}]{abgrall2021legend}
\bibinfo{author}{\bibfnamefont{N.}~\bibnamefont{Abgrall}},
  \bibinfo{author}{\bibfnamefont{I.}~\bibnamefont{Abt}},
  \bibinfo{author}{\bibfnamefont{M.}~\bibnamefont{Agostini}},
  \bibinfo{author}{\bibfnamefont{A.}~\bibnamefont{Alexander}},
  \bibinfo{author}{\bibfnamefont{C.}~\bibnamefont{Andreoiu}},
  \bibinfo{author}{\bibfnamefont{G.}~\bibnamefont{Araujo}},
  \bibinfo{author}{\bibfnamefont{F.}~\bibnamefont{Avignone~III}},
  \bibinfo{author}{\bibfnamefont{W.}~\bibnamefont{Bae}},
  \bibinfo{author}{\bibfnamefont{A.}~\bibnamefont{Bakalyarov}},
  \bibinfo{author}{\bibfnamefont{M.}~\bibnamefont{Balata}},
  \bibnamefont{et~al.}, \bibinfo{journal}{arXiv preprint arXiv:2107.11462}
  (\bibinfo{year}{2021}).

\end{thebibliography}

\end{document}